# Temporal Starvation in CSMA Wireless Networks

Cai Hong Kai, Soung Chang Liew
Department of Information Engineering, The Chinese University of Hong Kong
Email: {chkai6, soung}@ie.cuhk.edu.hk

*Abstract*— It is well known that links in CSMA wireless networks are prone to starvation. Prior works focused almost exclusively on *equilibrium starvation*. In this paper, we show that links in CSMA wireless networks are also susceptible to *temporal starvation*. Specifically, although some links have good equilibrium throughputs and do not suffer from equilibrium starvation, they can still have no throughput for extended periods from time to time. Given its impact on quality of service, it is important to understand and characterize temporal starvation. To this end, we develop a "trap theory" to analyze temporal throughput fluctuation. The trap theory serves two functions. First, it allows us to derive new mathematical results that shed light on the transient behavior of CSMA networks. For example, we show that the duration of a trap, during which some links receive no throughput, is insensitive to the distributions of the backoff countdown and transmission time (packet duration) in the CSMA protocol. Second, we can develop analytical tools for computing the "degrees of starvation" for CSMA networks to aid network design. For example, given a CSMA network, we can determine whether it suffers from starvation, and if so, which links will starve. Furthermore, the likelihood and durations of temporal starvation can also be computed. We believe that the ability to identify and characterize temporal starvation as established in this paper will serve as an important first step toward the design of effective remedies for it.

## I. INTRODUCTION

Starvation in communication networks is an undesirable phenomenon in which some users receive zero or close-to-zero throughputs. Wireless carrier-sense-multiple-access (CSMA) networks, such as Wi-Fi, are prone to starvation [1-5].

In CSMA networks, different stations compete with each other using the CSMA medium-access control (MAC) protocol. When a station hears its neighbors transmit, it will refrain from transmitting in order to avoid collisions.

If each station can hear all other stations, then the competition for airtime usage is fair. However, if each station hears only a subset of the other stations, and different stations hear different subsets of stations, then unfairness can arise.

The unfairness can be to the extent that some stations are totally starved while other stations enjoy good throughputs. As shown in our prior work [5], starvation can happen in many different CSMA network topologies, even in the absence of hidden terminals [19].

There are two types of starvation in CSMA wireless networks:
1) *Equilibrium Starvation* - A link could be starved because it receives near zero throughputs all the time.
2) *Temporal Starvation* - A link could also be starved in the *temporal* sense: it may have good long-term average throughput, but its throughput is near zero for excessively long periods from time to time.

This paper is devoted to a detailed quantitative study of temporal starvation. The study of equilibrium throughputs in many prior works [1-5] could only capture equilibrium starvation. The analysis of the temporal starvation is particularly challenging. To our knowledge, this paper is the first attempt to characterize temporal starvation analytically.

To characterize temporal starvation, we need to analyze the transient behavior of the underlying stochastic process. We emphasize that by "temporal", we do not mean that the starvation is temporary or ephemeral in nature. Indeed, temporal starvation in CSMA networks can be long-lasting.

Fig.1 shows an example of temporal starvation. We have a small grid network consisting of six links. All the links have good long-term average throughputs; yet they suffer from temporal starvation, as described below.

The carrier-sensing relationships among the links in the network are represented by the contention graph on the right of Fig. 1. In the contention graph, links are represented by vertices, and an edge joins two vertices if the transmitters of the two links can sense (hear) each other (i.e., the transmitters of the two links are within Carrier Sensing Range (CSRange) of each other). Thus, in this network, when links 1, 4, and 5 transmit, links 2, 3, and 6 cannot transmit, and vice versa.

The normalized equilibrium throughput of each link in the network can be shown to be around 0.5 either by simulation or by analysis using the method in [5]. However, as shown by the simulation results presented in Fig. 2, the temporal throughputs of links vary drastically over time.

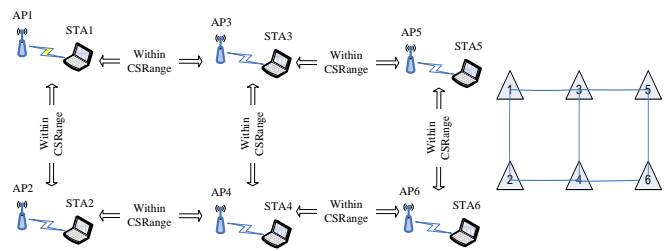

Fig.1. An example network and its associated contention graph.

Fig. 2 plots the normalized throughputs versus time for links 1 and 2. Each data point is the throughput averaged over a window of one second. As can be seen, once a link gets access to the channel, it can transmit consecutively for a long time; on the other hand, once it loses the channel, it also has to wait a long time before it has a chance to transmit again.

The above example is a small network. Temporal starvation can be more severe for larger networks. For example, in an $N \times M$ grid network similar to that in Fig. 1, but with larger $N$ and $M$, the active and idle periods are much longer than those shown in Fig. 2.

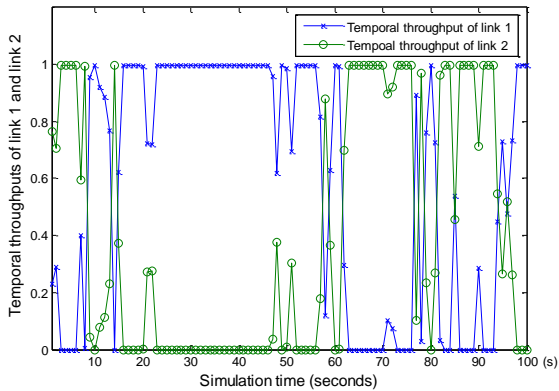

Fig.2. Temporal throughputs (averaged over time window of one second) of link 1 and link 2 in Fig. 1. Throughputs of other links exhibit similar fluctuations.

This paper is devoted to the identification and characterization of temporal starvation. Within this context, this paper has two contributions:

1. We propose a "trap theory" for the study of the temporal starvation in CSMA wireless networks, based on which a computational toolset for starvation characterization can be constructed.

2. We derive new analytical results related to the transient behavior of CSMA networks, providing new understanding to their transient behavior beyond the equilibrium analysis of prior works.

With respect to contribution 1, a trap is a subset of system states during which certain links receive zero or little throughputs; while the system evolves within the trap, these links suffer from temporal starvation. Based on the trap theory and the prior equilibrium analytical framework [5], we can construct computational tool to aid network design. For example, we can determine whether a given CSMA network suffers from starvation; if so, which links will starve, and whether the starvation is equilibrium or temporal in nature. Furthermore, for each link, the probability of temporal starvation and its duration can be characterized quantitatively. This ability to identify and characterize starvation is an important first step toward finding the remedies to circumvent it.

With respect to contribution 2, we show that the mean trap duration is insensitive to the distributions of countdown and transmission times, even if the backoff countdown process of the CSMA protocol is non-memoryless. We note that the 802.11 protocol is of this nature; hence, the practical relevance of this result. In addition to the insensitivity result, this paper establishes some asymptotic results to capture the dependencies of the trap duration on system parameters and network topology. Specifically, we show that the trap duration increases polynomially with the ratio of the mean transmission time to the mean backoff time in the CSMA protocol, and exponentially with the depth of the trap. Closed-form results for trap duration are derived for some regular networks.

**Related Work**

The focus of this paper is on temporal starvation in CSMA wireless networks. In particular, we are interested in networks in which the carrier sensing is "non-all-inclusive" [5] in that each link can only sense a subset of other links.

The equilibrium throughput of CSMA wireless networks has been well studied. Ref. [21] derived the equilibrium throughput of an "all-inclusive" network in which all links can sense all the other links. Ref. [2] and [5] investigated the non-all-inclusive case and showed that equilibrium throughputs of the links can be computed by modeling the network state as a time-reversible Markov chain. The temporal throughput fluctuations, however, were not considered.

Ref. [6][7][8] developed analytical models to evaluate the average transmission delay, delay jitter and the short-term unfairness in CSMA wireless network. However, they only considered the less interesting "all-inclusive" networks.

Ref. [9] considered two infinite CSMA networks with regular contention graphs: 1-D linear and 2-D grid networks. The border effects, fairness and phase transition phenomenon were investigated for both networks. Different from the regular networks studied in [9], this paper provides an analytical framework for characterizing temporal starvation in general CSMA wireless networks.

The remainder of this paper is organized as follows: Section II elaborates the motivations for our work and provides a qualitative overview of our approach. Section III introduces our system model and briefly reviews an equilibrium analysis. Section IV defines traps mathematically, presents a procedure to identify them and relate them to temporal starvation. Section V analyzes the duration of a trap captured by the ergodic sojourn time. The computational toolset for starvation characterization based on trap theory is constructed in Section VI. Section VII shows that the existing remedies for equilibrium starvation may not solve temporal starvation and remedies for temporal starvation are briefly discussed. Finally Section VIII concludes the paper.

## II. MOTIVATIONS AND QUALITATIVE OVERVIEW OF OUR APPROACH

An example of temporal starvation was given in the introduction. This section is devoted to a qualitative examination of the cause of the phenomenon, which sets the stage for the quantitative framework of our trap theory in Section IV.

For comparison and contrast with temporal starvation, let us first look at an example of equilibrium starvation. Consider a network with the contention graph shown in Fig. 3(a). Link 2 is sandwiched between links 1 and 3. When link 2 hears the transmission of either link 1 or link 3, its backoff countdown process will freeze. Fig. 3(b) shows that link 2 gets near-zero throughputs all the time, while the normalized throughputs of links 1 and 3 are close to 1 (the maximum throughput).

The diagram shown in Fig. 4 illustrates how the activities of link 1 and link 3 are sensed by the transmitter of link 2. Note that links 1 and 3 cannot hear each other and do not coordinate their transmissions. As far as link 2 is concerned, the transmissions of links 1 and 3 overlap randomly in time. Link 2 can only perform backoff countdown when both links 1 and 3 are idle, but the probability of this event is small because the backoff countdown period is typically much smaller than the transmission duration (e.g., in a typical 802.11b network, the mean transmission time is more than five times larger than the mean backoff time).

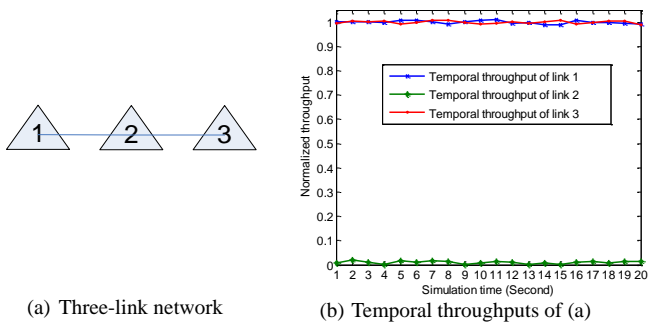

(a) Three-link network  (b) Temporal throughputs of (a)

Fig.3 Contention graph of a three-link network and the temporal throughputs (averaged over time window of one second) of links in the network.

The starvation of link 2 in Fig. 3 can be directly characterized by the equilibrium throughput. In fact, the normalized equilibrium throughputs of the three links can be shown to be approximately $[Th_1, Th_2, Th_3] = [1, 0, 1]$ using a quick back-of-the-envelop computation method proposed in [5]. Equilibrium starvation is characterized by near-zero equilibrium throughput. Ref. [5] has examined this issue in detail and much understanding about equilibrium starvation has been acquired.

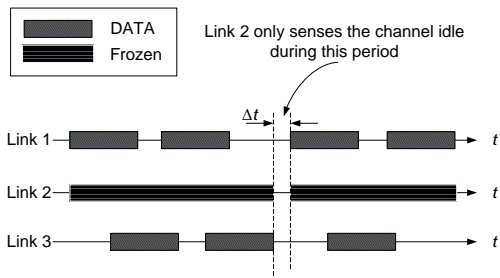

Fig. 4 The channel activities of link 1 and 3 sensed by the transmitter of link 2.

In general, starvation (equilibrium or temporal) occurs whenever a transmitter can sense more than one other links that cannot sense each other. Figuratively, the starved link is being "trapped" by the randomly overlapping transmission activities of its adjacent links. As will be seen later, temporal starvation of a link occurs when there are multiple "traps" in the network, some favorable and some unfavorable to the link.

Let us now look more closely into the temporal starvation of the network in Fig. 1(a). As mentioned in the introduction, the normalized long-term average throughput of each link in the network is around 0.5. Fig. 5(a) shows an example of a network in which the normalized equilibrium throughput of each link is also 0.5. That is, there is no difference between the long-term throughputs of the links of the networks shown in Fig. 1(a) and Fig. 5(a). However, unlike the network in Fig. 1, the network in Fig. 5 does not suffer from temporal starvation. As shown in Fig. 5(b), the temporal throughputs of the two links are constant around 0.5. This is quite different from the drastic throughput fluctuations of the links in Fig. 1(a) as shown in Fig. 2. In particular, links 1 and 2 in Fig. 5 compete with each other for the channel airtime without any "trap" phenomenon.

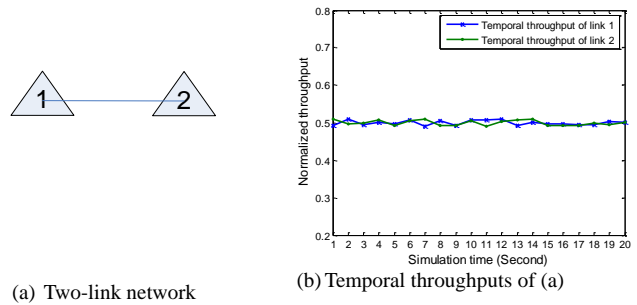

(a) Two-link network  (b) Temporal throughputs of (a)

Fig.5 Contention graph of a two-links network and the temporal throughputs (averaged over time window of one second) of links in the network.

To see the "trap" for the network in Fig. 1, we divide the six links into two groups: links $\{1,4,5\}$ and links $\{2,3,6\}$. The links in each group do not coordinate their transmissions in a direct manner under the random CSMA MAC protocol. Yet, they tend to have good (and bad) temporal throughputs at the same time. This is due to the "enemy-of-my-enemy-is-my friend" phenomenon, as described below.

Consider an instant when link 4 is transmitting. Since links $\{2,3,6\}$ can sense link 4, their backoff countdown will be frozen. Meanwhile, links 1 and 5 are free to perform their backoff countdown and transmit. When link 4 finishes its transmission, there is a good chance that links 1 and 5 are transmitting, and this freezes links $\{2,3,6\}$ as well. In this mode, links $\{1,4,5\}$ basically pre-empt links $\{2,3,6\}$ from transmitting. The backoff countdown of links $\{2,3,6\}$ advance very slowly. This continues until some time later when links $\{2,3,6\}$ take over to pre-empt the transmissions of links $\{1,4,5\}$. Thus, these two groups of links take turns to have good and bad throughputs. Their temporal starvations are caused by the "trap" set up by the other group.

Essentially, a link is trapped by a group of other adjacent links that do not coordinate their transmissions directly, but yet their activities are such that it is as if they work together to starve the link.

The above has described "traps" qualitatively. To characterize temporal starvation quantitatively, we need to define traps precisely. Only then can we measure the degree of temporal starvation and correlate it to the properties of traps. Specifically, the sojourn time of a trap and the first passage time between traps are related to the duration of a temporal starvation. The quantitative study of traps is the main focus of this paper. With the trap theory, we can then identify and quantitatively characterize temporal starvation in a general CSMA network.

## III. SYSTEM MODEL

In this section, we first present an idealized version of the CSMA network (ICN) to capture the main features of the CSMA protocol responsible for the interaction and dependency among links. The ICN model was used in several prior investigations [2][5][9]. The exact correspondence between the ICN model and the IEEE 802.11 protocol [10] can be found in our previous paper [5].

### A. The ICN model

In ICN, the carrier-sensing relationship among links is described by a contention graph as in many other prior papers [2] [5] [11]. Each link is modeled as a vertex. Edges, on the other hand, model the carrier-sensing relationships among links. There is an edge between two vertices if the transmitters of the two associated links can sense each other.

At any time, a link is in one of two possible states, active or idle. A link is active if there is a data transmission between its two end nodes. Thanks to carrier sensing, any two links that can hear each other will refrain from being active at the same time. A link sees the channel as idle if and only if none of its neighbors is active.

In ICN, each link maintains a *backoff timer*, $C$, the initial value of which is a random variable with an *arbitrary* distribution $f(t_{cd})$ and mean $E[t_{cd}]$. The timer value of the link decreases in a continuous manner with $dC/dt = -1$ as long as the link senses the channel as idle. If the channel is sensed busy (due to a neighbor transmitting), the countdown process is frozen and $dC/dt = 0$. When the channel becomes idle again, the countdown continues and $dC/dt = -1$ with $C$ initialized to the previous frozen value. When $C$ reaches 0, the link transmits a packet. The transmission duration is a random variable with *arbitrary* distribution $g(t_{tr})$ and mean $E[t_{tr}]$. After the transmission, the link resets $C$ to a new random value according to the distribution $f(t_{cd})$, and the process repeats. We define *the access intensity* of a link as the ratio of its mean transmission duration to its mean backoff time: $\rho = E[t_{tr}]/E[t_{cd}]$. In this paper, we will normalize time such that $E[t_{tr}] = 1$. That is, time is measured in units of average transmission duration. Thus, $\rho = 1/E[t_{cd}]$.

Let $x_i \in \{0,1\}$ denote the state of link $i$, where $x_i = 1$ if link $i$ is active (transmitting) and $x_i = 0$ if link $i$ is idle (actively counting down or frozen). The overall **system state** of ICN is $s = x_1 x_2 ... x_N$, where $N$ is the number of links in the network. Note that $x_i$ and $x_j$ cannot both be 1 at the same time if links $i$ and $j$ are neighbors because (i) they can sense each other; and (ii) the probability of them counting down to zero and transmitting together is 0 under ICN (because the backoff time is a continuous random variable).

The collection of feasible states corresponds to the collection of independent sets of the contention graph. An independent set (IS) of a graph is a subset of vertices such that no edge joins any two of them [12]. For a particular feasible state $x_1 x_2 ... x_N$, link $i$ is in the corresponding IS if and only if $x_i = 1$. Thus, we may also denote the system state by enumerating the active links in the state, i.e., $s = \{1,4,5\}$ represents a state in which links 1, 4 and 5 are active and the other links are idle. A maximal independent set (MaIS) is an IS that is not a subset of any other independent set [13], and a maximum independent set (MIS) is a largest maximal independent set [13]. Under an MaIS or an MIS, all non-active links are frozen, and none of them can become active.

As an example, Fig. 6 shows the state-transition diagram of the network in Fig. 1 under the ICN model. To avoid clutters, we have merged the two directional transitions between two states into one line. Each transition from left to right corresponds to the beginning of a transmission on one particular link, while the reverse transition corresponds to the ending of a transmission on the same link. For example, the transition from $\{1\}$ to $\{1,4\}$ is due to link 4's beginning to transmit; while the reverse transition from $\{1,4\}$ to $\{1\}$ is due to link 4's ending its transmission.

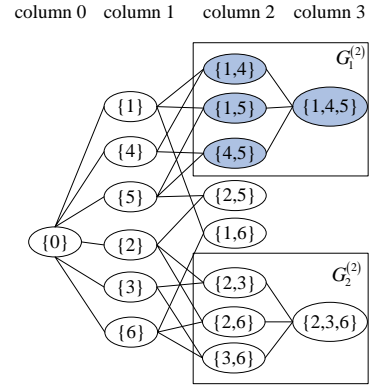

Fig. 6. The state-transition diagram of the network shown in Fig. 1. $G_1^{(2)}$ and $G_2^{(2)}$ are two traps.

### B. Equilibrium analysis

If we further assume that the backoff time and transmission time are exponentially distributed, then $s(t)$ is a time-reversible Markov process. For any pair of neighbor states in the continuous-time Markov chain, the transition from the left state to the right state occurs at rate $1/E[t_{cd}] = \rho$, and the transition from the right state to the left state occurs at rate $1/E[t_{tr}] = 1$. The stationary distribution of state $s$ can be computed by [5]

$$P_s = \frac{\rho^n}{Z} \ \forall \ s \in S^{(n)}, \text{ where } Z = \sum_n |S^{(n)}| \rho^n \quad (1)$$

In (1), $S^{(n)}$ is the subset of states with $n$ active links and $Z$ is the normalization factor. The fraction of time during which link $i$ transmits is $Th_i = \sum_{s:x_i=1} P_s$. We shall refer to $Th_i$ as the normalized throughput of link $i$.

For modest-size networks, [5] showed that we could take the limit $\rho \to \infty$ to accurately approximate the equilibrium normalized throughputs of links. For our example in Fig. 1, doing so gives

$$Th_1 = Th_2 = Th_5 = Th_6 \approx \lim_{\rho \to \infty} \frac{\rho + 3\rho^2 + \rho^3}{1 + 6\rho + 8\rho^2 + 2\rho^3} = 0.5$$
$$Th_3 = Th_4 \approx \lim_{\rho \to \infty} \frac{\rho + 2\rho^2 + \rho^3}{1 + 6\rho + 8\rho^2 + 2\rho^3} = 0.5 \qquad (2)$$

Furthermore, [5] showed that (1) is in fact quite general and does not require the system state $S(t)$ to be Markovian. In particular, (1) is insensitive to the distributions of the transmission time and the backoff time, given the ratio of their mean $\rho$. In other words, (1) still holds even if the backoff time and transmission time are not exponentially distributed.

## IV. Traps and Temporal Starvation

Section II has described traps qualitatively. A trap could occur when a link is surrounded by multiple links that cannot sense each other. In this section, we give the mathematical definition of traps and relate them to temporal starvation in CSMA wireless networks.

### A. Definition of Traps

A trap is a subset of "connected" states in which multiple links transmit and hog the channel for excessive time. While the system evolves within this subset of states, their neighboring links may get starved. Within the trap, the throughputs of these starved links may be much lower than their equilibrium throughputs. In this case, we say that temporal starvation occurs.

As an example, consider Fig. 1 again. States $\{\{1,4\},\{1,5\},\{4,5\},\{1,4,5\}\}$ and states $\{\{2,3\},\{2,6\},\{3,6\},\{2,3,6\}\}$ constitute two traps; links $\{2,3,6\}$ suffer from temporal starvation in the first trap, and links $\{1,4,5\}$ suffer from temporal starvation in the second trap.

Traps can be identified from the state-transition diagram of the ICN. Recall that for a non-MaIS state, the transition rate to a right neighbor state is $\rho$, and the transition rate to a left neighbor is 1. Note that the backoff countdown period is typically much smaller than the transmission duration in CSMA wireless networks (i.e., $\rho$ is large) and $\rho$ can be larger when TXOP is increased to reduce countdown overhead [22]. Large $\rho$ tends to push the system to states with more transmitting links. That is, in the state-transition diagram the movement from the right to the left is much more difficult than the movement from the left to the right. This could be seen from the relationship given in (1) as well, in which states with more transmitting links have higher probabilities through the factor $\rho^n$.

Before defining traps precisely, for illustration and motivation, let us look at the example of Fig.1 again. With respect to its state-transition diagram in Fig. 6, MIS $\{1,4,5\}$ and $\{2,3,6\}$ have the highest probabilities. Starting from either MIS, the process will next visit a state with one fewer transmitting link when it makes a transition. After that, the state may evolve back to the MIS (with rate $\rho$) or to a state with yet one additional idle link (with rate 1). However, large $\rho$ makes the movement to the left states a lot less likely. The system process tends to circulate among the subset of states composed of an MIS and its neighboring states. In particular, with large $\rho$, the system evolution will be anchored around the MIS, with departures from it soon drifting back to it. This will continue for a duration of time, depending on the "depth" of the trap (to be defined soon) and the value of $\rho$, until the system evolves to the other trap anchored by the other MIS.

To isolate the two traps anchored around the two MIS, we could truncate the left two columns of the state-transition diagram in Fig. 6. We could then define the sets of states connected to MIS $\{1,4,5\}$ and MIS $\{2,3,6\}$ as two traps, respectively (i.e., the states enclosed in the two boxes in Fig. 6). We could use a *transient* analysis to analyze the time it takes for the system to evolve out of a trap, which sheds light on the duration of temporal starvation.

Moving beyond the above illustrating example, we now present the exact definition of traps in a general CSMA network. Let us denote the graph corresponding to the complete state-transition diagram by $G$. In $G$, we arrange the states (vertices) such that the states with the same number of active links are in the same column. Label the column from left to right as $0, 1, 2, \cdots$ (i.e., the states in column $l$ have $l$ active links).

**Definition of the $l$-column truncated state-transition diagram:** The state-transition diagram with columns 0, 1, 2 and $l-1$ truncated, denoted by $G^{(l)}$, will be referred to as the $l$-column truncated state-transition diagram. Each state in the leftmost column of $G^{(l)}$ has $l$ transmitting links.

Note that when we truncate a state (vertex), we also eliminate the transitions (edges) out of it and into it. If two states are retained in a truncated graph, the transitions between them remain intact.

**Definition of state connectivity:** Two feasible states $s_i$ and $s_j$ are said to be connected if it is possible to find a path from $s_i$ to $s_j$ in the state transition diagram, and vice versa. Obviously, all the states are connected in $G$. This may not be the case, however, in $G^{(l)}$.

**Definition of disconnected subgraphs and traps in $G^{(l)}$:** $G^{(l)}$ may consist of a number of subgraphs: within each subgraph, all states are connected; the states between the subgraphs, however, are disconnected. Let $N_l$ denote the number of such disconnected subgraphs in $G^{(l)}$, and $G_1^{(l)}, G_2^{(l)}, \cdots, G_{N_l}^{(l)}$ denote the subgraphs themselves. A subgraph $G_i^{(l)}$ is said to be a *trap* if there are at least two columns in it.

The reason for requiring $G_i^{(l)}$ to have at least two columns to qualify as a trap is as follows. A general property of ICN is that in $G$, there is no direct transition (edge) between two states of the same column. Thus, if $G_i^{(l)}$ has only one column, then it must have only one single state; otherwise, the condi-tion that all states in $G_i^{(l)}$ are connected as defined above would not be fulfilled. This means that when the process enters $G_i^{(l)}$, with probability 1 the next state that the process will visit will be a state in the left of $G_i^{(l)}$. That is, regardless of $\rho$, the process will not get "trapped" in $G_i^{(l)}$ for long.

### Procedure to identify traps

As mentioned above, temporal starvation occurs within traps. To determine whether a given network suffers from temporal starvation, we need to study the traps in its state-transition diagram. We now describe a procedure to decompose the system states into traps in a hierarchical manner (in general, there could be traps within a trap). In practice, this procedure could be automated by a computer program as part of a toolset to identify and analyze temporal starvation for a given CSMA network contention graph.

We use the network on the left of Fig. 7 as an illustrating example. The state-transition diagram of the network is shown on the right of Fig. 7.

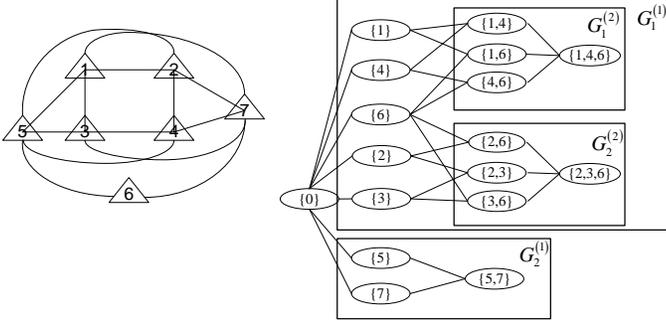

Fig. 7 An example network for illustrating the procedure to identify traps

1) **Step 1:** we find the minimum $l$ such that $G^{(l)}$ consists of at least two disconnected subgraphs. Among the subgraphs $G_1^{(l)}, G_2^{(l)}, \cdots, G_{N_l}^{(l)}$, if none of them has at least two columns, then there is no trap in the network. We call those $G_i^{(l)}$, $i \in \{1, ..., N_l\}$, with at least two columns the first-level traps. For the example of Fig. 7, the minimum $l = 1$, and $N_l = 2$. Both $G_1^{(1)}$ and $G_2^{(1)}$ are first-level traps as shown in Fig. 7(b).

**Definitions of roots and depth of a trap:** For each trap, we define the set of states with the maximum cardinality as the "roots" of the trap. Mathematically, the roots are

$$R\left(G_i^{(l)}\right) = \left\{s : |s| \geq |s'|, \forall s, s' \in G_i^{(l)}\right\}.$$

Furthermore, we define the depth of a trap as the cardinality of a root minus $l$:

$$D\left(G_i^{(l)}\right) = \max_{s \in G_i^{(l)}} |s| - l$$

In other words, $D\left(G_i^{(l)}\right) + 1$ is the number of columns in $G_i^{(l)}$; and for $G_i^{(l)}$ to qualify as a trap, $D\left(G_i^{(l)}\right) \geq 1$.

For easy reference, we write $Tr(s_r, l) = G_i^{(l)}$ (shortened as $Tr$) where $s_r$ is any one of the roots of the trap $G_i^{(l)}$ and $l$ is the index of the leftmost column of the trap.

Let $Th(i|Tr)$ denote the normalized throughput of link $i$ given that the process is within the trap $Tr$. Mathematically, $Th(i|Tr)$ is the conditional probability as follows:

$$\begin{aligned}Th(i|Tr) &= \Pr\{link\ i\ is\ active\,|\,the\ process\ is\ within\ Tr\} \\ &= \Pr\{s : x_i = 1 | s \in Tr\} = \sum_{s:x_i=1, s \in Tr} P_s / \sum_{s \in Tr} P_s \end{aligned} \quad (3)$$

where $P_s$ is given by (1).

We define the links which cannot receive a minimum targeted throughput while within the trap as the starving links of the trap, denoted by $S(Tr)$:

$$S(Tr) = \left\{i \mid Th(i|Tr) < \overline{Th}_{\text{temp}}\right\}$$

where $\overline{Th}_{\text{temp}}$ is determined by the requirement of the applications running on top of the wireless network.

For our example in Fig. 7, we have two first-level traps: $G_1^{(1)} \triangleq Tr(\{1,3,6\},1)$ (equivalently, $G_1^{(1)} \triangleq Tr(\{1,4,6\},1)$) and $G_2^{(1)} \triangleq Tr(\{5,7\},1)$. $R\left(G_1^{(1)}\right) = \{\{1,4,6\},\{1,3,6\}\}$ and $R\left(G_2^{(1)}\right) = \{\{5,7\}\}$; $D\left(G_1^{(1)}\right) = 2$ and $D\left(G_2^{(1)}\right) = 1$. For any $\overline{Th}_{\text{temp}} > 0$, we have $S\left(G_1^{(1)}\right) = $ links $\{5,7\}$ and $S\left(G_2^{(1)}\right) = $ links $\{1,2,3,4,6\}$.

2) **Step 2:** for each first-level trap, we increase $l$ further and check whether it can be further decomposed into a number of second-level traps.

For our example in Fig. 7, $G_2^{(1)}$ cannot be decomposed any further, while $G_1^{(1)}$ can be decomposed to two second-level traps: $G_1^{(2)} \triangleq Tr(\{1,4,6\},2)$ and $G_2^{(2)} \triangleq Tr(\{2,3,6\},2)$. $R\left(G_1^{(2)}\right) = \{\{1,4,6\}\}$ and $R\left(G_2^{(2)}\right) = \{\{2,3,6\}\}$; $D\left(G_1^{(2)}\right) = 1$ and $D\left(G_2^{(2)}\right) = 1$; for any $\overline{Th}_{\text{temp}} > 0$, we have $S\left(G_1^{(2)}\right) = $ links $\{2,3,5,7\}$ and $S\left(G_2^{(2)}\right) = $ link $\{1,4,5,7\}$.

3) **Further Steps:** Similarly, we construct the third-level traps by decomposing the second-level traps. Repeat this procedure until all the newly formed traps cannot be decomposed further.

*B. Definition of equilibrium and temporal starvation*

In this paper, we define equilibrium starvation as follows:

***Definition of equilibrium starvation:*** A link $i$ is said to suffer from equilibrium starvation if its equilibrium throughput is below a target reference throughput. That is,

$$Th_i < \overline{Th}_{\text{equil}} \quad (4)$$

for some $\overline{Th}_{\text{equil}} > 0$.

In this definition, $\overline{Th}_{\text{equil}}$ is determined by the requirement of the application running on top of the wireless network.

Equilibrium starvation can be identified directly from the equilibrium throughput, which has been well studied in prior works [1-6].

Next we use the trap technique to study the temporal starvation in CSMA networks. As will be argued in Section V, the depth of a trap $Tr$ is an important parameter characterizing the severity of the temporal starvation suffered by links $S(Tr)$. In particular, the duration of the temporal starvation grows exponentially with the depth. The following definition of temporal starvation is motivated by this result:

***Definition of temporal starvation:*** A link $i$ is said to suffer from temporal starvation if there is at least a trap $Tr$ with depth $D(Tr) \geq d_{\text{target}}$ in which link $i$ gets throughputs below $\overline{Th}_{\text{temp}}$ within the trap (i.e., link $i \in S(Tr)$).

Note that temporal starvation is defined with respect to two given parameters: (i) $\overline{Th}_{\text{temp}}$ is an application requirement; (ii) $d_{\text{target}}$ is determined not only by the application requirement, but also by the value of the access intensity $\rho$. Section V will study the duration of a trap as captured by the ergodic sojourn time and we will elaborate this definition further in Section VI.

## V. ANALYSIS OF TRAP DURATION

As demonstrated earlier, the duration of a trap is directly related to the severity of temporal starvation. We characterize the expected duration of a trap by its ergodic sojourn time and study its related properties. In particular, when $\rho$ is large, we obtain asymptotic analytical results for the computation of trap duration. In addition, the ergodic sojourn time of a trap is shown to be insensitive to the distributions of countdown and transmission times given their respective means: this implies our analysis on trap duration is applicable to general CSMA wireless networks, including 802.11 networks. Furthermore, an approximate computation method is proposed for the case when $\rho$ is not large and the approximation is validated by simulations. Finally, closed-form results for trap durations are derived for some regular networks.

### A. The expected first passage time from a state within a trap to the outside

We derive the expected trap duration by first focusing on the time it takes to exit the trap given that the system is in a particular state within the trap: $s \in Tr$. Specifically, we compute the expectation of the first passage time from a particular state $s$ within $Tr$ to the subset of the state space $B = G \setminus Tr$, denoted by $T(s \rightarrow B)$.

We index the columns of a particular trap $Tr = G_i^{(l)}$ with respect to the overall state-transition diagram $G$. That is, column $l$ refers to the leftmost column, and column $l+d$ refers to the rightmost column, of $Tr$, where $d$ is its depth. Let $A_k$, $l \leq k \leq l+d$ denote the states in column $k$ of the trap. We have the following theorem:

***Theorem 1:*** Consider a trap $Tr(s_r, l) = G_i^{(l)}$ within the state-transition diagram of a CSMA wireless network. For any state $s \in Tr(s_r, l)$ and $\rho > 1$, we have

$$T(s \rightarrow B) = \beta \rho^d + o(\rho^d), \quad (5)$$

where $d$ is the depth of the trap, and $\beta = \frac{|A_{l+d}|}{l|A_l|}$; $|A_l|$ is the number of states in the leftmost column of $Tr(s_r, l)$, and $|A_{l+d}|$ is the number of states in the rightmost column of $Tr(s_r, l)$.

***Proof:*** See Appendix A.

Theorem 1 indicates that starting with any state within the trap, the expected passage time to arrive at a state outside the trap is of order $\beta \rho^d$, where $\beta$ is a constant determined by the network topology and $d$ is the depth of the trap. Given a fixed network topology, $T(s \rightarrow B)$ increases polynomially with $\rho$. Given a fixed $\rho$, $T(s \rightarrow B)$ increases exponentially with $d$. We can see that for a large $\rho$ and a finite network, traps of higher depth are much more significant than traps of lower depth in terms of trap duration. Note that in (5) both $\beta$ and $d$ are both determined by the network contention graph and its state-transition diagram.

An interesting and significant observation of Theorem 1 is that for large $\rho$, the dominant term $\beta \rho^d$ in $T(s \rightarrow B)$ is independent of the state $s$. Different states yield different $T(s \rightarrow B)$ only through the term $o(\rho^d)$. This means that the duration of the trap depends only weakly on where the journey into the trap begins. We will make the notion of the duration of traps more concrete below.

### B. Ergodic sojourn time of a trap

Theorem 1 is related to the expected remaining trap duration given that the system is currently in a particular state within a trap. We now study the ergodic sojourn time of a trap, which provides a measure of the expected duration of the trap.

All visits to a trap $Tr$ begin at some state within it. Assuming the system process is ergodic, we would like to derive the probability of a visit to $Tr$ beginning at state $s \in Tr$. Let $h_{Bs}$ be the average number of visits to $Tr$ per unit time that begins at state $s \in Tr$, defined as follows:

$$\begin{aligned} h_{Bs} &= \lim_{t \to \infty} \frac{1}{t}\left[\text{the number of transitions from } B \text{ to } s \text{ in } (0,t)\right] \\ &= \sum_{s' \in B} P_{s'} \upsilon_{s's} \end{aligned}, \quad (6)$$

where $\upsilon_{s's}$ is the transition rate from state $s'$ to state $s$ in the complete continuous-time Markov chain $S(t)$.

Given the fact that the system just arrives at the trap, we specify the initial distribution as

$$P_s(0) = \frac{h_{Bs}}{\sum_{s' \in Tr} h_{Bs'}}, \quad s \in Tr \quad (7)$$

$$P_s(0) = 0, \quad s \in B$$

***Definition of Ergodic sojourn time of a trap:*** the ergodic sojourn time of a trap $Tr$ is defined as the time for the system process to evolve out of the trap given that the initial condition is specified by (7):

$$T_V(Tr) = \sum_{s \in Tr} P_s(0) T(s \to B). \quad (8)$$

In fact, a similar definition is used in [14] to characterize the expected sojourn time of visits to a group of states in general Markov chains.

According to our definition of traps, when the process evolves into a trap $Tr = G_i^{(l)}$ from subset $B$, the state it arrives at must be in column $l$ of the overall state-transition diagram. Equation (1) indicates that the states in the same column of the state-transition diagram have equal stationary probability. In particular, $P_{s'}$ in (6) are the same for different $s' \in B$ for which there is a valid transition $s' \to s$. Furthermore, for each state $s$ in the leftmost column of $Tr$, we have $\sum_{s' \in B} \upsilon_{s's} = l\rho$. Then, (7) can be rewritten as

$$P_s(0) = \frac{h_{Bs}}{\sum_{s' \in Tr} h_{Bs'}} = \frac{l\rho}{|A_l| l\rho} = \frac{1}{|A_l|} \text{ if } s \text{ is in column } l \text{ of } Tr; \quad (9)$$

$$P_s(0) = 0, \quad \text{otherwise.}$$

In other words, the journey into the trap begins at any of the $|A_l|$ states in the leftmost column of the trap with equal probability. Thus, (8) can be written as

$$T_V(Tr) = \sum_{s \in Tr, |s|=l} \frac{T(s \to B)}{|A_l|}. \quad (10)$$

Combining Theorem 1 and (10), the ergodic sojourn time $T_V(Tr)$ satisfies

$$T_V(Tr) = \beta \rho^d + o(\rho^d) \quad (11)$$

For large $\rho$, $T_V(Tr)$ is dominated by the term $\beta \rho^d$. For moderate $\rho$, we will provide a simple method to approximate $T_V(Tr)$ in Part D.

According to our definition of traps and temporal starvation, the ergodic sojourn time of a trap provides a lower bound for the duration of temporal starvation. Once the system process evolves into a trap, on average the starving links of the trap will receive near-zero throughputs for at least the duration of the trap. This explains why we define temporal starvation in terms of the depth of the trap in Section IV-B. Furthermore, the starving links may starve much longer if the system returns to the trap without passing through states in which the links do not starve, in which case the expected passage time among traps becomes important to characterize the duration of temporal starvation. We will provide further details on this issue in Section VI.

*C. Dependency of Ergodic sojourn time on distributions of the backoff time and transmission time*

As established in [5], the equilibrium throughput of a CSMA wireless network is insensitive to the distributions of the backoff and transmission times given their means, even if the backoff process is one that has memory as it alternates between active countdown and frozen state due to transmissions by neighbors. Here, we show that the ergodic sojourn time of a trap also has this insensitivity property.

***Theorem 2:*** The ergodic sojourn time of a trap, $T_V(Tr)$, is insensitive to the distributions of countdown and transmission times given their respective means $E[t_{cd}]$ and $E[t_{tr}]$, even if the backoff countdown process of the CSMA protocol is non-memoryless (e.g., countdown continues with the previously frozen value after emerging from a frozen state when the neighbors stop transmitting, as in 802.11 networks).

***Proof:*** See Appendix B.

The insensitivity of the ergodic sojourn time means that our analysis of $T_V(Tr)$ is applicable to a general CSMA network in which the backoff time and the transmission are not exponentially distributed and the backoff process actually has memory (e.g., 802.11 networks). Hence, our treatment on temporal starvation is applicable to a general CSMA network.

*D. Approximation of $T_V(Tr)$ when $\rho$ is not large*

The asymptotic results in (11) are applicable to the large $\rho$ case. For a given $\rho$, whether the highest-order term of $\rho$ dominates is also dependent on the network topology.

Here, we consider a finer approximation of $T_V(Tr)$. In principle, we could use equation (6.2.5) of [14] to obtain a closed-form expression of $T(s \to B)$. However, the computation is of high complexity. We show that we could construct a simpler Markov chain that is a birth-death process to approximate $T_V(Tr)$. As can be seen later, $T_V(Tr)$ can be computed in closed-form easily in the simplified Markov chain.

Let $A_{l-1} \subseteq B$ denote the subset of states in $B$ which are directly connected to states in $A_l$ of $Tr$ (Note that the states in $B$ that are directly connected to $Tr$ must be in column $l-1$ in the complete state-transition diagram $G$). First, we aggregate the states in $A_i$ into a single state, denoted by $\tilde{s}_i$ in the simplified Markov chain. We do this for all columns $i$, $l-1 \leq i \leq l+d$. Thus, the state $\tilde{s}_i$ is the union of all states in $A_i$. We see that each column in the original Markov chain is encapsulated into a single state in the simplified Markov chain. At any given time $t$, by definition $P_{\tilde{s}_i}(t) \triangleq \sum_{s_i \in A_i} P_{s_i}(t)$. Just as transitions could only occur between states of adjacent columns in the original Markov chain, transitions could only occur between adjacent states in the simplified Markov chain.

Next, we need to determine the transition rate between two adjacent states in the simplified Markov chain. We do this by equating the probability fluxes in the two Markov chains, as follows.

In the original Markov chain, the total probability flux from all the states encapsulated in $\tilde{s}_i$ to all the states encapsulated in $\tilde{s}_{i-1}$ on the left is given by $\sum_{s_i \in A_i} P_{s_i}(t) \cdot i = i \sum_{s_i \in A_i} P_{s_i}(t)$. For the simplified Markov chain, we define the effective transition rate $\mu(\tilde{s}_i \to \tilde{s}_{i-1})$ from $\tilde{s}_i$ to $\tilde{s}_{i-1}$ according to this probability flux. Specifically, $P_{\tilde{s}_i}(t)\mu(\tilde{s}_i \to \tilde{s}_{i-1}) \triangleq i \sum_{s_i \in A_i} P_{s_i}(t)$. Since by definition $P_{\tilde{s}_i}(t) \triangleq \sum_{s_i \in A_i} P_{s_i}(t)$, we have $\mu(\tilde{s}_i \to \tilde{s}_{i-1}) = i$.

The transition rate from $\tilde{s}_i$ to $\tilde{s}_{i+1}$, $\mu(\tilde{s}_i \to \tilde{s}_{i+1})$, is more tricky. Although it is true that stationary probabilities $P_{s_i}$ for all $s_i$ in the same column are equal, that is not the case when the system is in transience. That is, $P_{s_i}(t)$ in the definition of $P_{\tilde{s}_i}(t) \triangleq \sum_{s_i \in A_i} P_{s_i}(t)$ may not be equal during transience. This is where approximation is made in our simplified Markov chain. Specifically, for our approximation, we assume $P_{s_i}(t)$ for all $s_i \in A_i$ are equal. Note that this is at least true at the beginning of the visit to the trap (i.e, at $t=0$) according to (9). As time progresses, during the system evolution within the trap, this is in general not strictly true and is only an approximation. As demonstrated in Appendix C, only when the states in the same column of the trap have the same number of right neighbors, the above approximation is exact (Lemma 10).

The probability flux of the union state $\tilde{s}_i$ to the right is $P_{\tilde{s}_i}(t)\mu(\tilde{s}_i \to \tilde{s}_{i+1}) \triangleq \sum_{s_i \in A_i} P_{s_i}(t) \cdot \rho n_{s_i}$, where $n_{s_i}$ is the number possible right transitions from state $s_i$ in the original Markov chain. With our above approximation, we have $P_{\tilde{s}_i}(t) \triangleq \sum_{s_i \in A_i} P_{s_i}(t) \approx |A_i| P_{s_i}(t)$ where $P_{s_i}(t)$ are equal for all $s_i \in A_i$. Thus, $\mu(\tilde{s}_i \to \tilde{s}_{i+1}) \approx \frac{1}{|A_i|}\sum_{s_i \in A_i} \rho n_{s_i} = \frac{|A_{i+1}|}{|A_i|}(i+1)\rho$ (Note that $\sum_{s_i \in A_i} n_{s_i} = |A_{i+1}|(i+1)$ because this is the total number of the edges linking states in $A_{i+1}$ and states in $A_i$ in the original Markov chain). In summary, we have the following for our simplified Markov chain:

$$\begin{aligned}\mu(\tilde{s}_i \to \tilde{s}_{i-1}) &= i \\ \mu(\tilde{s}_i \to \tilde{s}_{i+1}) &= \frac{|A_{i+1}|}{|A_i|}(i+1)\rho\end{aligned} \quad (12)$$

where $l-1 < i < l+d$.

With the transition rates in (12), $\tilde{s}_{l-1}, \tilde{s}_l, \tilde{s}_{l+1}, \cdots, \tilde{s}_{l+d}$ form a birth-death process. The expected first passage time from $\tilde{s}_l, \cdots, \tilde{s}_{l+d}$ to $\tilde{s}_{l-1}$, $T(\tilde{s}_i \to \tilde{s}_{l-1})$, $l \leq i \leq l+d$ can be computed by [Section 5.2, 14]

$$T(\tilde{s}_i \to \tilde{s}_{l-1}) = \sum_{k=0}^{d}\left\{\sum_{j=0}^{\min(k,i-l)}\frac{|A_{l+d-k+j}|}{(l+j)|A_{l+j}|}\right\}\rho^{d-k} \quad (13)$$

The ergodic sojourn time of the trap can be computed as $T(\tilde{s}_l \to \tilde{s}_{l-1})$. That is,

$$\begin{aligned}T_V(Tr) = T(\tilde{s}_l \to \tilde{s}_{l-1}) &= \sum_{k=0}^{d}\frac{|A_{l+d-k}|}{l|A_l|}\rho^{d-k} \\ &= \sum_{k=0}^{d}\frac{|A_{l+d-k}|}{|A_{l+d}|}\beta\rho^{d-k}\end{aligned}. \quad (14)$$

*E. Ergodic sojourn time of Regular networks*

We examine several regular networks and see how the trap theory helps the understanding of the temporal performance of CSMA wireless networks. Although in general (14) is an approximation of the ergodic sojourn time of traps, it is exact for the following specific networks. That is, (14) yields exact closed-form results for trap durations derived below. This is because for these networks, the states in the same column of a trap have the same number of right neighbors (see Lemma 10).

*1) Ring network:* Consider a 1-D ring network with $N$ links. Label the links as $1, 2, 3, \cdots, N$.

i) When $N$ is odd, write $N = 2L+1$. Each MIS has $L$ active links (i.e., the right-most column is column $L$) and all the MIS get connected through column $L-1$. We cannot find an $l$ such that $N_l \geq 2$ and at least one of the unconnected subgraphs in $G^{(l)}$ qualify as a trap. That is, there is no trap in the network and hence no temporal starvation.

ii) When $N$ is even, write $N = 2L$. In $G^{(L-1)}$ there are two traps composed of states in which links $\{1,3,5,\cdots,2L-1\}$ and links $\{2,4,6,\cdots,2L\}$ take turn to hog the channel. The depth of both traps, however, is only 1.

According to Theorem 1 and Lemma 10 in Appendix C, the ergodic sojourn time of both traps can be computed as

$$T_V(Tr) = \frac{\rho}{L(L-1)} + \frac{1}{L-1} \quad (15)$$

As in (15), given a fixed $L$, $T_V(Tr)$ increases linearly with $\rho$. The larger the $\rho$, the more severe the temporal starvation. Fig. 8 shows the temporal throughputs of two typical links of the two traps with respect to different $\rho$. In the simulation, we fixed $N=4, L=2$ and varied the value of $\rho$. The throughputs of both links are measured over every $T=50$ ms. The typical access intensity in 802.11 networks, $\rho_0$ is 5.35. In the three sets of simulations, we set $\rho = 2\rho_0$, $\rho = 4\rho_0$ and $\rho = 8\rho_0$, respectively. From Fig. 8, we can see that as $\rho$ increases, the temporal starvation becomes more severe.

Equation (15) indicates that $T_V(Tr)$ is roughly inversely proportional to $L(L-1)$. Given a fixed $\rho$, the sojourn time of the trap decreases quickly with $L$. In particular, $N=4, L=2$ is the worst case. Fig.9 shows the temporal throughputs of two typical links of the two traps measured over successive 50-ms interval. We set $\rho = 8\rho_0$, and $N=4, 8$ and 16. As can be seen, temporal starvation is the most severe when $N=4$, and gradually disappears as $N$ increases.

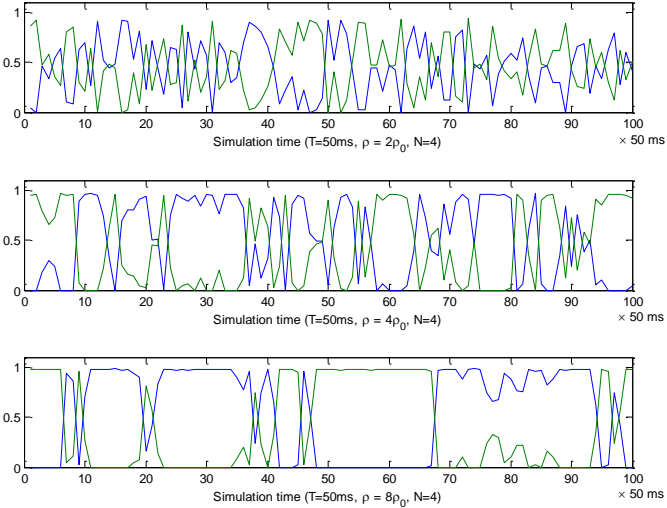

Fig. 8 Throughputs of two typical links, one in each of the two traps of an N=4 ring network, measured over successive 50-ms intervals, for $\rho = 2\rho_0, 4\rho_0$, and $8\rho_0$.

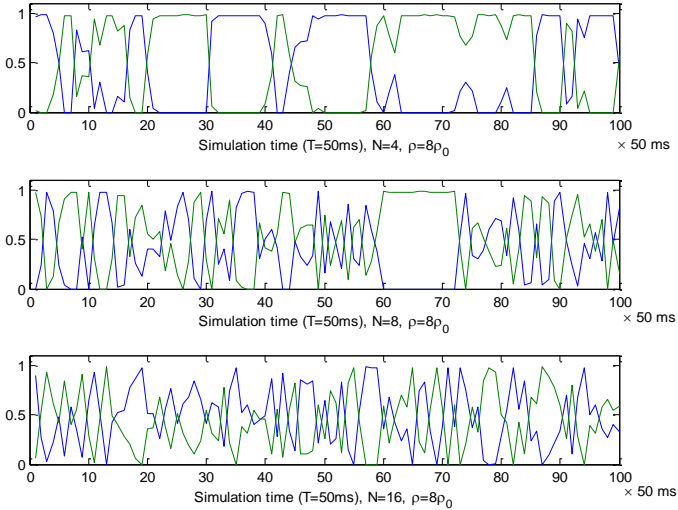

Fig.9. Throughputs of two typical links, one in each of the two traps of ring networks with N=4, 8, and 16, and $\rho = 8\rho_0$ measured over successive 50-ms intervals.

*2) Linear network:* Consider a linear network with $N$ links, label the links as $1, 2, 3, \cdots, N$.

i) When $N$ is even, write $N = 2L$. Each MIS has $L$ active links. It is not difficult to verify that all the MISs get connected through column $L-1$. According to our specification of traps, there is no trap in the network;

ii) When $N$ is odd, write $N = 2L+1$. In $G^{(L)}$ there is a trap composed of states in which links $\{1, 3, 5, \cdots, 2L+1\}$ keep transmitting. The depth of the trap is 1.

According to Theorem 1 and Lemma 10 in Appendix C, the ergodic sojourn time of the trap is

$$T_V(Tr) = \frac{\rho}{L(L+1)} + \frac{1}{L} \qquad (16)$$

Comparing (16) to (15), we can see that the linear network has temporal performance similar to the ring network. In particular, when $N$ is fixed, links $\{2, 4, \cdots, 2L\}$ suffer from more severe starvation as $\rho$ increases. When $\rho$ is fixed, the worst case appears when $L=1$. That is, it is a three-link linear network in which link 2 gets starved as shown in Fig. 3.

*3) Network of Fig.1:* As demonstrated in Section IV-A, there are two traps. The two traps have the same ergodic sojourn time due to symmetry.

Invoking (14), the ergodic sojourn time of both traps is

$$T_V(Tr) = \frac{\rho}{6} + \frac{1}{2} \qquad (17)$$

Note that in the network of Fig. 1, each state in the same column of the trap has the same number of right neighbors. That is, the computation of (17) is actually exact and no approximation is made. However, this is not generally true for all the 2-D grid networks.

*F. Simulation results to validate our approximate computation of $T_V(Tr)$*

For general networks, (14) is only an approximation. Next we use an ICN simulator to examine the accuracy of it. The ICN simulator is implemented using MATLAB programs. We generate ten 20-link random networks in which each link has on average three neighbors. In each simulation run, we gather the statistics of $T_V(Tr)$ and compare them with computations using (14).

Define $\Delta T_V$ as the ratio between prediction error and the simulated ergodic sojourn time of the particular trap. That is, $\Delta T_V = (T_V - \tilde{T}_V)/\tilde{T}_V$ in which $T_V$ is our approximate ergodic sojourn time by (14) and $\tilde{T}_V$ is the simulated ergodic sojourn time. Table I lists $\Delta T_V$ in the ten 20-link random networks for $\rho = 10\rho_0$. Averaging over ten networks, we find that approximation (14) can achieve accuracy within 0.24% error.

To further motivate our approximation of (14), we also compare the ergodic sojourn time computed by the highest-order term of $\rho$ only (i.e., $\beta \rho^d$ in (11)) to the simulated value. Define $\Delta T'_V = (\beta \rho^d - \tilde{T}_V)/\tilde{T}_V$ where $\beta \rho^d$ is defined in (11). As can be seen in Table I, on average it underestimates by 4.15%. The approximation of (14) is closer to the simulated ergodic sojourn time.

Another question is how well (14) and the term $\beta \rho^d$ approximate under different access intensity, $\rho$. In the following, we examine the accuracy of (14) for different $\rho$ in a 20-link network in which each link has on average three neighbors. As shown in Table II, The error of $\beta\rho^d$ decreases with $\rho$; this is not surprising because the $\beta\rho^d$ is the asymptotic result for large $\rho$. Interestingly, the error of (14) remains small for even for small $\rho$.

Table II. $\Delta T_V$ for networks of different value of $\rho$

| $\rho/\rho_0$ | 5 | 10 | 15 | 20 |
|---|---|---|---|---|
| $\Delta T_V$ | 0.87% | -0.13% | -0.08% | -0.88% |
| $\Delta T_V'$ | -10.79% | -6.26% | -3.09% | -1.96% |

Table I. $\Delta T_V$ in ten 20-link random networks for $\rho=10\rho_0$

| Network Number | #1 | #2 | #3 | #4 | #5 | #6 | #7 | #8 | #9 | #10 | Average |
|---|---|---|---|---|---|---|---|---|---|---|---|
| $\Delta T_V$ | -1.57% | 0.32% | 0.94% | 0.32% | -0.53% | 0.33% | 1.83% | -0.54% | 0.74% | 0.53% | 0.24% |
| $\Delta T_V'$ | -4.65% | -3.74% | -4.73% | -4.77% | -4.04% | -3.27% | -3.56% | -4.34% | -3.55% | -4.84% | -4.15% |

## VI. ANALYZING TEMPORAL STARVATION USING TRAP THEORY

This section is devoted to analyze temporal starvation using the trap theory. Specifically, we propose the procedure to identify temporal starvation from traps and list the corresponding starving links. Besides the expected trap duration studied in Section V, the severity of temporal starvation is further characterized by the probability of traps and the passage times among traps. One potential outcome of our analysis above is to construct a computational toolset to quantitatively characterize the temporal starvation phenomenon in a general CSMA wireless network.

### A. Procedure to identify temporal starvation

Given the procedure to identify traps in the network, the procedure to identify temporal starvation is quite straightforward:

First, all the traps in the network are identified using the procedure described in Section IV-A. As can be seen in (11), the ergodic sojourn time of a trap increases polynomially with $\rho$. Given the same QoS requirement (e.g., the longest tolerant delay), we can determine $d_{\text{target}}$ with respect to $\rho$. That is, we set $d_{\text{target}}$ larger for a small $\rho$ and smaller when $\rho$ is large. We then go through all the traps with depth no less than $d_{\text{target}}$ and identify the links that suffer from temporal starvation. Incidentally, if $d_{\text{target}} > 1$ (i.e., the number of columns is more than 2), we could also modify the procedure in Section IV-A by directly redefining the definition of traps to require them to at least $d_{\text{target}} + 1$ rather than just two columns.

Let us illustrate the procedure with the example in Fig. 7. For simplicity, let us assume $d_{\text{target}} = 1$. For any $\overline{Th}_{\text{temp}} > 0$, links 1 and 4 suffer from temporal starvation in both trap $G_2^{(1)}$ and $G_2^{(2)}$; links 2 and 3, in both traps $G_2^{(1)}$ and $G_1^{(2)}$; links 5 and 7, in traps $G_1^{(1)}$, $G_1^{(2)}$ and $G_2^{(2)}$; link 6, in trap $G_2^{(1)}$.

For the above example, all links suffer from temporal starvation. However, their *probabilities* and *durations* of temporal starvation can be quite different. In general, the *significance* of a trap $Tr$ in terms of inducing temporal starvation on links $S(Tr)$ depends on two of its properties: the probability of $Tr$ and the duration of $Tr$. The duration of a trap has been carefully analyzed in Section V. We next study the probability of a trap.

### B. Probability of traps

We define the probability of a trap as the stationary probability for the process to be within the trap:

$$\Pr\{Tr\} = \sum_{s \in Tr} P_s \qquad (18)$$

The probability of a trap $Tr$ characterizes how likely the links in $S(Tr)$ will suffer from temporal starvation because of $Tr$.

The probability of a trap can be directly obtained from the time-reversible Markov chain described in Section III-B. For our example in Fig. 7, we have

$$\begin{aligned}\Pr\{G_1^{(1)}\} &= \frac{5\rho + 6\rho^2 + 2\rho^3}{1 + 7\rho + 7\rho^2 + 2\rho^3} \\ \Pr\{G_2^{(1)}\} &= \frac{2\rho + \rho^2}{1 + 7\rho + 7\rho^2 + 2\rho^3} \\ \Pr\{G_1^{(2)}\} &= \frac{3\rho^2 + \rho^3}{1 + 7\rho + 7\rho^2 + 2\rho^3} \\ \Pr\{G_2^{(2)}\} &= \frac{3\rho^2 + \rho^3}{1 + 7\rho + 7\rho^2 + 2\rho^3}\end{aligned} \qquad (19)$$

Given that the typical value of $\rho$ in 802.11 networks, $\rho_0$, is 5.35, we have $\Pr\{G_1^{(1)}\} = 92.61\%$, $\Pr\{G_2^{(1)}\} = 7.21\%$, $\Pr\{G_1^{(2)}\} = 43.58\%$ and $\Pr\{G_2^{(2)}\} = 43.58\%$. That is, for any $Th_{\text{temp}} > 0$, links 1, 2, 3 and 4 suffer from temporal starvation with probability $7.21\% + 43.58\% = 50.79\%$; links 5 and 7 with probability $92.61\%$ and link 6 with probability $7.21\%$.

### C. The expected first passage time from a trap to another trap

Recall that the sojourn time of a trap only provides a lower bound for the duration of temporal starvation. The duration of starvation also depends on how many times the process will revisit the trap before it finally visits states in which the starving links enjoy good throughputs. It is possible for the system to exit and enter the same trap repeatedly before it finally enters another trap. In this case, it is important to analyze the first passage time between traps to characterize the severity of temporal starvation. In Appendix D we mathematically define the expected first passage time from a trap to another trap and design computation methods to approximate the expected first passage time. We overview the procedure below.

Suppose that we want to compute the expected first passage time between two traps $Tr_i$ and $Tr_j$, denoted by $T_P(Tr_i \to Tr_j)$. In the complete state-transition diagram $G$ we find all the traps that have no intersections with $Tr_i$ or $Tr_j$ while making sure that each state in $G$ is included in at most one trap. Denote the set of these traps together with $Tr_i$ and $Tr_j$ by $\mathbf{Tr} = \{Tr_1, Tr_2, \cdots, Tr_n\}$.

Next we construct a simplified stochastic process $\hat{S}(t)$ to compute the passage time between traps. We aggregate all the states within a trap $Tr_i$ into a single state, denoted by $s_{Tr_i}$. We do this for all the traps within $\mathbf{Tr}$ and transform the complete state-transition diagram $G$ to a simplified state-transition diagram $G^*$. For trap states $s_{Tr_i}$ in $G^*$, we define the rate the process $S(t)$ leaves the state as $\upsilon_{s_{Tr_i}} = 1/T_V(Tr_i)$, where $T_V(Tr_i)$ can be approximately computed by (14). Furthermore, we assume that the ergodic sojourn time of traps, the countdown and transmission times are exponentially distributed. Then $S(t)$ is a continuous-time Markov chain. The expected passage time between two trap-states can be computed using a standard technique in general Markov chains. More details can be found in Appendix D.

*D. Temporal Analysis of General CSMA Networks*

For general CSMA networks, with theory and tools developed thus far, we can construct an analytical toolset to study the temporal behavior of throughputs. The tool can be implemented by a computer program for modest-size CSMA networks. The inputs to the program are the network topology in the form of a contention graph and the value of $\rho$. The outputs of the program are as follows: 1) the list of starving links; 2) the list of traps in the network; 3) the probability of traps, 4) the durations of traps and 5) the expected first passage time between traps.

Refer to our example in Fig.7, the user inputs the contention graph shown on the left of Fig.7 and the value of $\rho$. As on the right of the Fig.7, the computer program produces the Markov chain together with identification of traps using the procedure described in Section IV-*A*, upon which we obtain the lists of starving links and traps in the network. Then the user may want to find out the likelihood of links starvation and the durations of such starvations as follows.

All the links in Fig.7 suffer from temporal starvation. Equation (19) characterizes the probabilities of traps in the network, from which we can compute the probabilities of the occurrence of temporal starvation for each link.

Invoking (14), the ergodic sojourn time of traps can be computed as

$$T_V(G_1^{(1)}) = \frac{\rho^2}{5} + \frac{6}{5}\rho + 1$$

$$T_V(G_2^{(1)}) = \frac{\rho}{2} + 1 \qquad (20)$$

$$T_V(G_1^{(2)}) = T_V(G_2^{(2)}) = \frac{\rho}{6} + \frac{1}{2} + \frac{4}{1+2\rho} + \frac{2}{1+4\rho}$$

Using the approximate computation method proposed in Appendix *D*, we compute the expected first passage time between traps as follows.

$$T_P(G_1^{(1)} \to G_2^{(1)}) = \frac{7}{10}\rho^2 + \frac{21}{5}\rho + \frac{7}{2} + \frac{1}{2\rho}$$

$$T_P(G_2^{(1)} \to G_1^{(1)}) = \frac{7}{10}\rho + \frac{7}{5} + \frac{1}{5\rho} \qquad (21)$$

$$T_P(G_1^{(2)} \to G_2^{(2)}) = T_P(G_2^{(2)} \to G_1^{(2)}) = \rho + 3 + \frac{4}{1+2\rho} + \frac{2}{1+4\rho}$$

When $\rho$ is large, the expected duration of traps becomes rather large and hence the temporal starvation becomes more severe. In our example, links 1, 2, 3, 4 and 6 have good equilibrium throughputs; however, they still get starved when the system process evolves into trap $G_2^{(1)}$. To see this, we set up a simulation in which we initialized by letting links 5 and 7 transmit first (i.e., the system process starts within the trap $G_2^{(1)}$). Fig. 10 shows the temporal throughputs measured over successive 50-ms intervals. As can be seen, at the beginning period ($0 \sim 250$ ms), links 5 and 7 have the maximum throughputs while the other links receive zero throughputs, since the system process is within trap $G_2^{(1)}$. After that, the process evolves to trap $G_2^{(2)}$, in which links 1, 4, 5 and 7 get starved while links 2, 3 and 6 enjoy good throughputs. After $80T$ in the figure, the process evolves into trap $G_1^{(2)}$, in which links 2, 3, 5 and 7 starve. In the simulation, we observed that as time evolves, the system process transits among the three traps and all the links take turns to suffer from temporal starvation.

In the network of Fig.7, links 5 and 7 are the most prone to temporal starvation and get starved in both traps $G_1^{(2)}$ and $G_2^{(2)}$. Most of the time links $\{1,4\}$ and links $\{2,3\}$ alternate to receive good and zero throughputs. Link 6, however, have good throughputs in both traps $G_1^{(2)}$ and $G_2^{(2)}$. Finally, links 1, 2, 3, 4 and 6 may get starved in trap $G_1^{(2)}$, although the probability of this starvation is small. The probability of link starvation and the expected duration of temporal starvation (i.e., the expected duration of traps and the expected passage time from a trap to another) can be computed by (19), (20) and (21), respectively.

In general, our work allows the design of an automated computational tool to identify and quantitatively characterize starvation phenomenon in CSMA wireless networks. Given the state-transition diagram of the system, it is easy to determine computationally whether the truncated diagram $G^{(l)}$ is connected and then identify traps [15]. Hence, the complexity mainly relies in generating the state space of the system process as described in Section III-*B*. For modest-size CSMA wireless networks, we can quickly identify temporal starvation using our toolset described above. The complexity issue of large CSMA wireless networks will be tackled in our future studies.

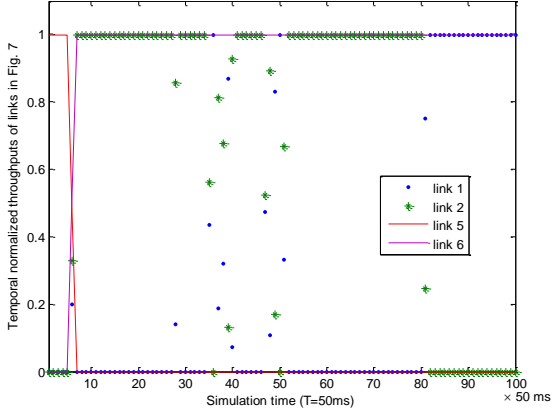

Fig.10. Throughputs of links 1, 2, 5 and 6 in the network of Fig. 7, and $\rho = 100\rho_0$ measured over successive 50-ms intervals. Link 3, link 4 and link 7 have similar throughputs to that of link 2, link 1 and link 5, respectively.

## VII. REMEDIES FOR TEMPORAL STARVATION

We show that the existing remedies designed for solving equilibrium starvation may not work well as far as temporal starvation is considered. Meanwhile, possible remedies are briefly discussed with details left for future investigations.

### A. Remedies for Equilibrium Starvation may not solve Temporal Starvation

Various methods have been proposed in the literature to alleviate starvation [16-18] in CSMA networks. However, most of them focused on mitigating equilibrium starvation and rarely considered temporal performance. We next argue that the remedies which can solve equilibrium starvation may not work well as far as temporal performance is considered.

*1) Channel assignment schemes:* Multiple channels are widely adopted to boost link throughputs [16]. Consider the network shown in Fig. 11 (a). It is an eight-link CSMA wireless network. If all the links in the network use the same channel, it is not difficult to verify that each link obtains a normalized throughput of 0.25 for large $\rho$ as shown in [5]. Next we assume that we have two orthogonal channels $f_1$, $f_2$ to assign. Fig. 11(b) and Fig. 11(c) show two possible channel-assignment configurations. These two configurations result in the same equilibrium throughput for each link (i.e., 0.5 for each link for large $\rho$). That is, there is no difference between the two configurations if they are evaluated in terms of equilibrium performance. However, when temporal performance is taken into account, Configuration 1 in Fig. 11(b) is obviously better than Configuration 2 in Fig. 11(c). The reason is that in Configuration 2 there are traps in both channels (i.e., the contention graph is a four-links ring network for both channels). Links under Configuration 2 may suffer from temporal starvation. As shown in (17) and Fig. 8, when $\rho$ is large, the links in the network will take turns to suffer from temporal starvation.

If temporal starvation is to be avoided, the channel assignment problem should be formulated with it in mind. In particular, it will be desirable to design the channel assignment algorithm to remove traps with large depth. The existing channel assignment schemes proposed so far, however, have not considered this.

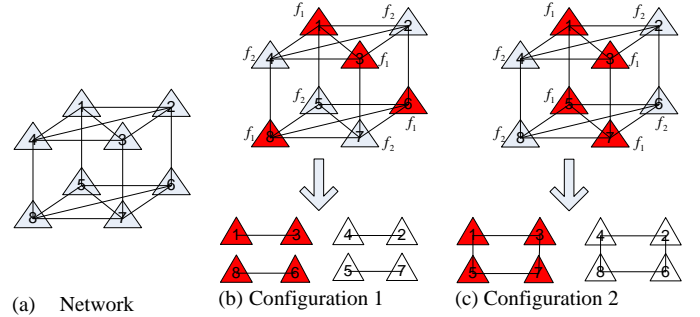

(a) Network  (b) Configuration 1  (c) Configuration 2
Fig. 11 An illustrating example showing that existing channel assignment schemes may not solve temporal starvation.

*2) Adaptive CSMA schemes:* Recent works in [17, 18] introduced an elegant adaptive CSMA scheduling algorithm that can achieve maximal system utility in a distributive manner. They optimize the aggregate user utility $U(Th_1, Th_2, ..., Th_N)$ by adjusting $\rho_i$, $i = 1,...,N$ of different links under the ICN model.

Although the adaptive CSMA algorithm proposed in [17, 18] can maximize system utility defined in terms of link equilibrium throughputs, we observed that temporal starvation could still exist in the adaptive CSMA networks. Take an $8*8$ grid network as an example, our simulation results indicate that regardless of whether the adaptive CSMA algorithm converges or not, temporal starvation still exists. We implemented Algorithm 2 of [17] and use the same parameters as defined in [17]. That is, $\partial = 0.23$, $\beta = 3$ and the utility function of each flow is defined by $\upsilon_m(f_m) = \log(f_m + 0.01)$. As can be seen in Fig. 12, the temporal throughputs of links in the network are always alternating between 0-1 over each 2.5 seconds. Temporal starvation has not been removed by the adaptive CSMA algorithm.

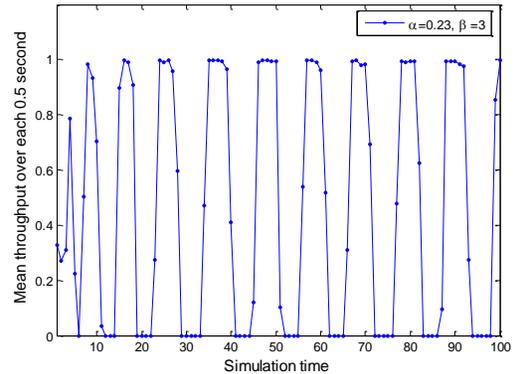

Fig.12. Throughputs of link 1 in an $8*8$ grid network measured over successive 0.5s intervals when Algorithm 2 of [17] is implemented. Throughputs of other links exhibit similar fluctuations.

### B. Remedies of Temporal Starvation

We have seen that remedies for equilibrium starvation or solutions that can allow links to achieve acceptable equilibrium throughputs may still suffer from temporal starvation. With the foundation established in this paper, the next challenge is to design remedies for temporal starvation. We briefly discuss two possible approaches here, relegating the details to future work.

*1) Remove "traps" in the network:* Recall from Section IV that temporal starvation occurs within traps. Thus, one direct way to eliminate temporal starvation is to remove traps, which may be done through multiple channel assignment. Refer to our example in Fig. 1, an immediate approach to remove temporal starvation is that assigning the links $\{1,4,5\}$ to channel $f_1$ and the links $\{2,3,6\}$ to another channel $f_2$.

In general CSMA wireless networks, we can try to assign the active links of an MaIS to the same channel and each MaIS to different channels (e.g., in Fig. 1, links $\{1,4,5\}$ to channel $f_1$ and links $\{2,3,6\}$ to channel $f_2$). It is possible that the available channels are not enough to accommodate all MaIS, in which case we could assign channels to the MaISs with the most active links. That is, we try to maximize $|\text{MaIS}_1| \cup |\text{MaIS}_2| \cup \cdots |\text{MaIS}_n|$ where $n$ is the number of available channels.

The disadvantage of this method is that the complete contention graph is needed to perform channel assignment. Also outstanding is an effective method to judge/predict whether traps exist in the network for a given channel assignment before the channels of the links are fixed. Although for small networks like Fig. 11 (a), we can obtain the appropriate channel configuration quickly even by hand computation, how to deal with larger networks are challenging and left for future study.

*2) "Unsaturated" adaptive CSMA schemes:* Previously we showed that directly applying adaptive CSMA schemes proposed in [17] does not remove temporal starvation. However, in our further investigation, we have worked out an "unsaturated" adaptive CSMA scheme which can eliminate temporal starvation at the cost of a small penalty in the system utility achieved. The details will be presented elsewhere due to the limited space of this paper.

## VIII. CONCLUSION

This paper has proposed a framework called the trap theory for the study of temporal starvation in CSMA networks. The theory serves two functions: 1) it allows us to establish analytical results that provide insights on the dependencies of transient behavior of CSMA networks on the system parameters (e.g., how does access intensity $\rho$ affects temporal starvation); 2) it allows us to build computational tools to aid network design (e.g., a computer program can be written to determine whether a given CSMA network suffers from starvation, the degree of starvation, and the links that will be starved). In both regards, we have demonstrated the accuracy of our results by extensive simulations.

A goal of this paper is to enrich our understanding on starvation phenomenon in CSMA wireless networks. We show that equilibrium throughput analysis is not enough to capture all the potential starvations and the network enhancement solutions that do not take temporal throughput fluctuations into account may need to be reevaluated and further optimized. For example, we show that the adaptive CSMA algorithm which can achieve the maximal system utility [17] does not remove temporal starvation in some networks.

Throughout this paper, we have separated the treatment of temporal starvation from that of equilibrium starvation. One may wonder whether it is possible to characterize both kinds of starvation using a single definition. In Appendix *E* we propose a possible approach. Furthermore, we provide a sufficient condition in terms of a special set of traps to judge where a particular link will starve according to our united definition.

REFERENCES

[1] P. C. Ng, S. C. Liew, "Throughput Analysis of IEEE802.11 Multi-hop Ad-hoc Networks," *IEEE/ACM Trans. Networking*, vol. 15, no. 2, June 2007.
[2] X. Wang and K. Kar, "Throughput Modeling and Fairness Issues in CSMA/CA Based Ad hoc networks," *IEEE INFOCOM*, Miami, 2005.
[3] M. Garetto, et al., "Modeling Per-flow Throughput and Capturing Starvation in CSMA Multi-hop Wireless Networks," *IEEE INFOCOM* 2006.
[4] K. Medepalli, F. A. Tobagi, "Towards Performance Modeling of IEEE 802.11 Based Wireless Networks: A Unified Framework and Its Applications," *IEEE INFOCOM* 2006.
[5] S. Liew, C. Kai, J. Leung and B. Wong, "Back-of-the-Envelope Computation of Throughput Distributions in CSMA Wireless Networks", *to appear in IEEE Transactions on Mobile Computing*, 2010. Technical report also available at http://arxiv.org/pdf/0712.1854.
[6] C.E. Koksal, H. Kassab, H. Balakrishnan, "An analysis of short-term fairness in wireless media access protocols," in: *Proceedings of the ACM SIGMETRICS*, 2000.
[7] Z.F. Li, S. Nandi, A.K. Gupta, "Modeling the Short-term Unfairness of IEEE 802.11 in Presence of Hidden Terminals," *Elsevier, performance evaluation* 63 (2006) 441-462.
[8] Marcelo M. Carvalho, J. J. Garcia-Luna-Aceves, Delay Analysis of IEEE 802.11 in Single-Hop Networks, *Proceedings of the 11th IEEE International Conference on Network Protocols,* p.146, November 04-07, 2003.
[9] M. Durvy, O. Dousse, P. Thiran, "Border Effects, Fairness, and Phase Transition in Large Wireless Networks", *IEEE Infocom 2008*, Phoenix, USA.
[10] IEEE Std 802.11-1997, IEEE 802.11 Wireless LAN Medium Access Control (MAC) and Physical Layer (PHY) Specifications.
[11] F. Zuyuan, and B. Bensaou, "Fair bandwidth sharing algorithms based on game theory frameworks for wireless ad-hoc networks," *IEEE INFOCOM 2004*, vol. 2, pp. 1284-1295.
[12] Independent set, http://en.wikipedia.org/wiki/Independent_set_(graph_theory).
[13] Maximal independent set, http://en.wikipedia.org/wiki/Maximal_independent_set.
[14] J. Keilson, Markov Chain Models-Rarity and Exponentiality, Springer-Verlag, 1979.
[15] Graph connectivity, http://en.wikipedia.org/wiki/Graph_connectivity.
[16] A. Mishra. V. Shrivastava, D. Agrawal, S. Banerjee and S. Ganguly, "Distributed Channal Management in Uncoordinated Wireless Environments," *Mobicom 2006*, p.170-181, Los Angeles, California, USA.
[17] L. Jiang, J. Walrand, "A Distributed CSMA Algorithm for Throughput and Utility Maximization in Wireless Networks," to appear in *IEEE/ACM Transactions on Networking*, 2010.
[18] M. Chen, S. Liew, Z. Shao and C. Kai, "Markov Approximation for Combinatorial Network Optimization," *in IEEE Infocom*, 2010.
[19] L. B. Jiang, S. C. Liew, "Hidden-node Removal and Its Application in Cellular WiFi Networks," *IEEE Trans. Vehicular Tech.*, vol. 56, no. 5, Sep. 2007.
[20] S. Liew, Y. Zhang and D. Chen, "Bounded Mean-Delay Throughput and Non-Starvation Conditions in Aloha Network," in *IEEE Transition on Networking*, Vol.17, Issue 5, pp.1606-1618, 2009.

[21] G. Bianchi, "Performance Analysis of the IEEE 802.11 Distributed Coordination Function," *IEEE Journal on Selected Areas in Communications*, vol. 18, no. 3, Mar. 2000.
[22] IEEE Std 802.11e-2005, IEEE 802.11 Wireless LAN Medium Access Control (MAC) and Physical Layer (PHY) Specifications: Medium Access Control (MAC) Quality of Service Enhancements.
[23] Invertible matrix, http://en.wikipedia.org/wiki/Invertible_matrix.

## APPENDIX A: PROOF OF THE THEOREM 1

For simplicity, we write $Tr$ in place of $Tr(s_r, l)$. The outline of the proof is as follows:

**Step 1:** We uniformize the continuous-time Markov chain $S(t)$ to a discrete Markov chain with one-step transition matrix $\prod_{\upsilon_0}$, where $\upsilon_0$ is the rate used to regulate the rates at which the process leaves a state [14]. Then the substochastic process within a trap $Tr$ is characterized by a submatrix $\prod_{\upsilon_0, Tr}$ of $\prod_{\upsilon_0}$. Let $\lambda_1$ denote the maximal eigenvalue of $\prod_{\upsilon_0, Tr}$, and let $\vec{V}_1 = (v_{s_1}, v_{s_2}, \cdots, v_{s_i}, \cdots)$ denote the associated normalized eigenvector with $\sum_{s_i \in Tr} v_{s_i} = 1$ and $v_{s_i} > 0\ \forall s_i \in Tr$.

**Step 2:** Set the initial probability distribution to $\vec{V}_1$ (i.e., $\vec{P}_0 = \vec{V}_1$). We show that the expected time for the process to evolve out of the trap, $T_{E|\vec{V}_1}$, satisfies

$$T_{E|\vec{V}_1} \triangleq \sum_{s_i \in Tr} v_{s_i} T(s_i \to B) = \frac{1}{\upsilon_0} \frac{1}{1-\lambda_1}. \quad (A1)$$

**Step 3:** We show that $\lambda_1$ satisfies

$$\frac{1}{\upsilon_0} \frac{1}{1-\lambda_1} = \beta \rho^d + o(\rho^d) \quad (A2)$$

where $\beta$ is a constant. Combining with (A1), we have

$$T_{E|\vec{V}_1} = \beta \rho^d + o(\rho^d). \quad (A3)$$

**Step 4:** To compute $\beta$, we construct a simple Birth-Death process $S_{Tr^a}(t)$. Let $\prod^a$ be the one-step transition matrix of the uniformized Markov chain of $S_{Tr^a}(t)$. We show that $\lambda_1$, the eigenvalue of $\prod_{\upsilon_0, Tr}$, is also an eigenvalue of $\prod^a$. Then the expected exit time of $S_{Tr^a}(t)$ can be written as

$$T_{E|\vec{V}_1^a} = \frac{1}{\upsilon_0} \frac{1}{1-\lambda_1} \quad (A4)$$

where $\vec{V}_1^a$ is the associated eigenvector of $\prod^a$. On the other hand, for Birth-Death processes we have a closed-form expression of the expected passage times. Making use of the results in Section 5.1 of [14], we show that

$$T_{E|\vec{V}_1^a} = \frac{|A_{l+d}|}{l|A_l|} \rho^d + o(\rho^d) \quad (A5)$$

Combining (A3), (A4) and (A5), we obtain $\beta = \frac{|A_{l+d}|}{l|A_l|}$, where $l$ is the index of the leftmost column of the trap $Tr$, $|A_l|$ is the number of states in the leftmost column of $Tr$, and $|A_{l+d}|$ is the number of states in the rightmost column of $Tr$. Thus,

$$T_{E|\vec{V}_1} = \frac{|A_{l+d}|}{l|A_l|} \rho^d + o(\rho^d) \quad (A6)$$

**Step 5:** We show that $T(s \to B) = \beta \rho^d + o(\rho^d)\ \ \forall s \in Tr$. That is, the highest-order term of $T(s \to B)$, $\beta \rho^d$, is the same as the highest-order term of $T_{E|\vec{V}_1}$.

The above steps are detailed below:

### 1. Markov Chain Uniformization

Consider the system process $S(t)$ introduced in Section III-*A*. Let $\upsilon_{s_i}$ be the rate at which the process makes a transition when in state $s_i$, and $\upsilon_{s_i s_j}$ be the transition rate from state $s_i$ to state $s_j$. Thus, $\upsilon_{s_i} = \sum_{s_j} \upsilon_{s_i s_j}$.

For a finite network, $S(t)$ is finite and hence "uniformizable". Let $\upsilon_0 \geq \sup \upsilon_{s_i}$, $\upsilon_0 < \infty$ (e.g., we could let $\upsilon_0 \triangleq N \max(\rho, 1)$, where $N$ is the number of links in the network). We can construct the associated discrete time process $S_k^*$ for $S(t)$. The continuous-time Markov process $S(t)$ is transformed to a discrete time process $S_k^*$ with the following one-step transition matrix

$$\prod_{\upsilon_0} = (p_{s_i s_j}), \text{ where }$$
$$\begin{cases} p_{s_i s_j} = \upsilon_{s_i s_j}/\upsilon_0 & \text{for } i \neq j \\ p_{s_i s_i} = 1 - \upsilon_{s_i}/\upsilon_0 \end{cases} \quad (A7)$$

On average, there is a transition event for $S_k^*$ every $1/\upsilon_0$ time units. The epochs at which transitions occur in $S_k^*$ are a Poisson process $K_{\upsilon_0}(t)$ with rate $\upsilon_0$.

**Substochastic process of $S(t)$: $S_{Tr}(t)$**

Consider a trap $Tr$ with column $l$ being the leftmost column. We partition the state space of $S(t)$ into two sets, $Tr$ and $B = G \setminus Tr$. For transient analysis, we look at the substochastic process within the trap $S_{Tr}(t)$. We write $\prod_{\upsilon_0, Tr}$ for the submatrix of $\prod_{\upsilon_0}$ on $Tr$. Note that since the states in column $l$ can transit to the states which are not included in the trap, the submatrix $\prod_{\upsilon_0 Tr}$ is a substochastic matrix. For the rest of the proof, we deal with the discrete-time Markov chain associated with $\prod_{\upsilon_0, Tr}$.

Let $\lambda_1$ and $\vec{V}_1$ denote the maximal eigenvalue and the associated normalized eigenvector of $\prod_{\upsilon_0 Tr}$, respectively. By normalization, we mean $\vec{V}_1 \cdot \vec{1}^T = 1$. From Perron–Frobenius theorem (See Section 1.2 of [14]), we have that $1 > \lambda_1 > 0$, $\vec{V}_1 > 0$ (i.e., all elements in $\vec{V}_1$ are positive) and $\vec{V}_1 \Pi_{v_0 Tr} = \lambda_1 \vec{V}_1$.

## 2. To prove (A1)

Set the initial condition to be $\vec{P}_0 = \vec{V}_1$. After the $k^{th}$ transition, the system probability becomes $\vec{P}_k = \vec{V}_1 \Pi_{v_0 Tr}^k = \lambda_1^k \vec{V}_1$. Note that $\lambda_1^k \vec{V}_1 \vec{1}^T = \lambda_1^k$ is the probability that the system is still in the trap after $k$ transitions. Let $M$ be the number of transitions it takes the discrete Markov process to exit the trap. Then, $E[M] = \sum_{k=0}^{\infty} \Pr\{M > k\} = \sum_{k=0}^{\infty} \lambda_1^k = 1/(1-\lambda_1)$. Eqn. (A1) is obtained by noting that each of these transitions takes an average of $1/v_0$ time units.

## 3. To prove (A2)

To prove (A2), we first prove Lemma 1 below. For a pair of neighboring states, $s_i$ and $s_j$ in the discrete Markov chain, the probability flow from $s_i$ to $s_j$ in one transition is

$$flow(s_i \to s_j) = P_{s_i} p_{s_i s_j} = \begin{cases} P_{s_i} \rho / v_0 & \text{if } |s_j| - |s_i| = 1 \\ P_{s_i} / v_0 & \text{if } |s_j| - |s_i| = -1 \end{cases}.$$

For conciseness, the above expression can be rewritten as

$$flow(s_i \to s_j) = P_{s_i} \rho^{\frac{1}{2}(|s_j| - |s_i| + 1)} / v_0 \qquad (A8)$$

Set the initial condition as $\vec{P}_0 = \vec{V}_1 = (v_{s_1}, v_{s_2}, \cdots v_{s_i}, \cdots)$. Let us define $\sigma_{s_i} = v_{s_i} \rho^{-|s_i|}$ as the "potential" of state $s_i$. Thus, the net probability gain of $s_i$ from $s_j$ after the first transition is

$$gain(s_i \leftarrow s_j) = flow(s_j \to s_i) - flow(s_i \to s_j)$$
$$= v_{s_j} \rho^{\frac{1}{2}(|s_i| - |s_j| + 1)} / v_0 - v_{s_i} \rho^{\frac{1}{2}(|s_j| - |s_i| + 1)} / v_0.$$
$$= \rho^{\frac{1}{2}(|s_i| + |s_j| + 1)} (\sigma_{s_j} - \sigma_{s_i}) / v_0$$

Note that $gain(s_i \leftarrow s_j) > 0$ if and only if $\sigma_{s_j} > \sigma_{s_i}$. In other words, there is a net probability flow from the high potential state to the low potential state.

**Lemma 1:** With initial condition $\vec{P}_0 = \vec{V}_1$, $\forall s_i, s_j \in Tr$, the probability gain after the first transition, $gain(s_j \leftarrow s_i)$, satisfies $|gain(s_j \leftarrow s_i)| \triangleq |v_{s_i} p_{s_i s_j} - v_{s_j} p_{s_j s_i}| \leq 1 - \lambda_1$.

**Proof:** We prove the lemma by contradiction. Write $\delta = 1 - \lambda_1$. Suppose that there exist $s_i, s_j \in Tr$ such that $|gain(s_j \leftarrow s_i)| > \delta$. That is, $gain(s_j \leftarrow s_i) > \delta$ or $gain(s_j \leftarrow s_i) < -\delta$. Since the latter can be transformed to the former by exchanging $s_i$ and $s_j$, we only need to consider $gain(s_j \leftarrow s_i) > \delta$ in the contradiction proof below.

We partition the state space of $Tr$ into two disjoint subsets, $I$ and $J$. Let $I$ be the subset of states in $Tr$ whose potentials are no less than that of $s_i$. That is, $I = \{s \in Tr : \sigma_s \geq \sigma_{s_i}\}$ and define $J = Tr \setminus I$. Note that $s_j \in J$. The potentials of the states in $J$ are less than the potentials of the states in $I$. This means $gain(s_j \leftarrow s_i) > 0$, $\forall s_i \in I, s_j \in J$. Then,

$$gain(J \leftarrow I) \triangleq \sum_{s \in I} \sum_{s' \in J} gain(s' \leftarrow s) \geq gain(s_j \leftarrow s_i) > \delta$$
$$\Rightarrow gain(I \leftarrow J) < -\delta$$

Unless proven otherwise, $I$ could contain states in column $l$. That means $I$ could also potentially lose probability to states outside $Tr$. That is,

$$gain(I) = gain(I \leftarrow J) - (\text{possible probability loss from } I \text{ to the outside})$$
$$\leq gain(I \leftarrow J) < -\delta$$

(A9).

On the other hand, we have $\vec{V}_1 \Pi_{v_0, Tr} = \lambda_1 \vec{V}_1$. After the first transition, each state $s_i$ within $Tr$ loses a probability of $(1 - \lambda_1) v_{s_i} = v_{s_i} \delta$. Then, $gain(I) = -\sum_{s_i \in I} v_{s_i} \delta > -\delta$, which contradicts (A9). □

Next we prove Lemma 2 (i.e., (A2)) below.

**Lemma 2:** The maximal eigenvalue of $\Pi_{v_0 Tr}$, $\lambda_1$, satisfies $\frac{1}{v_0} \frac{1}{1-\lambda_1} = \beta \rho^d + o(\rho^d)$, where $\beta$ is a constant. Its associated eigenvector $\vec{V}_1 = (v_{s_1}, v_{s_2}, \cdots, v_{s_i}, \cdots)$ satisfies $v_{s_i} = \frac{c}{\rho^m} + o\left(\frac{1}{\rho^m}\right)$, where $m = l + d - |s_i|$ and $c$ is a constant such that $\lim_{\rho \to \infty} c = \frac{1}{|A_{l+d}|}$, with $|A_{l+d}|$ being the number of states in the right-most column of the trap.

**Proof:** Invoking (A1), we have $(1-\lambda_1) = \frac{1}{v_0 T_{E|\vec{V}_1}}$. It is easy to see that the exit time $T_{E|\vec{V}_1}$ should be an increasing function of $\rho$ because for larger $\rho$, left transitions are less likely than right transitions in the trap. Therefore, $(1-\lambda_1)$ should be an increasing function of $1/\rho$. Then we could write $\delta = 1 - \lambda_1$

$$= \left(\frac{c'}{\rho^e} + o\left(\frac{1}{\rho^e}\right)\right) \frac{1}{v_0} \text{ where } \left(\frac{c'}{\rho^e} + o\left(\frac{1}{\rho^e}\right)\right) \text{ is the series expansion of } (1-\lambda_1). \text{ Both } c' \text{ and } e \text{ are unknown. In the following we prove that } e = d \text{ by contradiction}$$

Set the initial condition to $\vec{P}_0 = \vec{V}_1$. After the first transition, the system has a probability leakage of $\delta$, and each state $s_i$ within the trap loses a probability of $v_{s_i}\delta$. According to our definition of traps, we know that the overall probability of the trap is lost via column $l$. We can write

$$\sum_{|s_i|=l} P_{s_i} \frac{l}{\upsilon_0} = \delta \Rightarrow \sum_{|s_i|=l} P_{s_i} = \frac{c'}{l}\frac{1}{\rho^e} + o\left(\frac{1}{\rho^e}\right) \Rightarrow \underset{|s_i|=l}{Sup}\{P_{s_i}\} \propto \frac{1}{\rho^e}$$

Note that in the above, $l/\upsilon_0$ is the transition probability from $s_i$ to its left neighbors (i.e., the outside of the trap). Let $s_a$ be any of $\arg\underset{|s_i|=l}{sup}\{P_{s_i}\}$. Then, $v_{s_a} = \underset{|s_i|=l}{Sup}\{P_{s_i}\}$ $= \frac{c}{\rho^e} + o\left(\frac{1}{\rho^e}\right), |s_a| = l$. Since it has at least one right state, let $s_b$ denote any of its right states. Invoking Lemma 1, we have

$$\left|\frac{v_{s_a}\rho}{\upsilon_0} - \frac{v_{s_b}}{\upsilon_0}\right| \le \delta \Rightarrow \frac{1}{\upsilon_0}\left|v_{s_a}\rho - v_{s_b}\right| \le \delta$$

$$\Rightarrow v_{s_a}\rho - \upsilon_0\delta \le v_{s_b} \le v_{s_a}\rho + \upsilon_0\delta$$

$$\Rightarrow \frac{c}{\rho^{e-1}} + o\left(\frac{1}{\rho^{e-1}}\right) \le v_{s_b} \le \frac{c}{\rho^{e-1}} + o\left(\frac{1}{\rho^{e-1}}\right)$$

$$\Rightarrow v_{s_b} = \frac{c}{\rho^{e-1}} + o\left(\frac{1}{\rho^{e-1}}\right)$$

For any left state of $s_b$, denoted by $s_c$, we have

$$\frac{1}{\upsilon_0}\left|v_{s_c}\rho - v_{s_b}\right| \le \delta \Rightarrow \frac{v_{s_b}}{\rho} - \frac{\delta\upsilon_0}{\rho} \le v_{s_c} \le \frac{v_{s_b}}{\rho} + \frac{\delta\upsilon_0}{\rho}$$

$$\Rightarrow v_{s_c} = \frac{c}{\rho^e} + o\left(\frac{1}{\rho^e}\right)$$

For any right state of $s_b$, denoted by $s_d$, similarly we have $v_{s_d} = \frac{c}{\rho^{e-2}} + o\left(\frac{1}{\rho^{e-2}}\right)$. Since all states are connected in the trap, we have $v_{s_i} = c\rho^{|s_i|-l-e} + o\left(\rho^{|s_i|-l-e}\right)$ where $c$ is a constant.

Thus for $|s_i| = l+d$, $v_{s_i} = c\rho^{d-e} + o\left(\rho^{d-e}\right)$. If $e < d$, then there exists a value $\rho < +\infty$ such that $v_{s_i} > 1$; If $e > d$, then there exists a value $\rho < +\infty$, such that $\sum_{s_i \in Tr} v_{s_i} < 1$. Thus, $e$ must be equal to $d$, $\delta = 1 - \lambda_1 \sim 1/(\rho^d \upsilon_0)$, and $v_{s_i}|_{|s_i|=l+d} = c + o(1)$. We have $\sum_{s_i \in Tr} v_{s_i} = 1 \Rightarrow |A_{l+d}|c + o(1) = 1$

$$\Rightarrow \lim_{\rho \to \infty} c = \frac{1}{|A_{l+d}|}.$$

Hence, $(1-\lambda_1)\upsilon_0 = \upsilon_0 \delta = \frac{1}{\beta \rho^d} + o\left(\frac{1}{\rho^d}\right)$ and $v_{s_i} = \frac{c}{\rho^m}$ $+ o\left(\frac{1}{\rho^m}\right)$, $i = 1,\cdots,n$ where $\beta$ and $c$ are constants and $m = l + d - |s_i|$. □

Thus far, we have proved (A2) and (A3).

### 4. To find $\beta$

In the following, we first construct a new discrete-time Markov chain. We then show that the eigenvalue associated with the "trap" of the new discrete-time Markov chain is also $\lambda_1$. After that, we show that the new discrete-time Markov chain is the uniformized Markov chain of a birth-death process. From there, we make use of a result in Chapter 5 of [14] to derive that $\beta = \frac{|A_{l+d}|}{l|A_l|}$.

1) Construction of a new discrete-time Markov chain

Consider the discrete-time Markov chain of the original substochastic process, which is characterized by $\prod_{\upsilon_0 Tr}$. Let $A_i$, $l \le i \le l+d$, denote the set of states in column $i$ of $Tr$. We aggregate the states in $A_i$ for each column $i$ into a single state, denoted by $\tilde{s}_i$, in a new discrete-time Markov chain (i.e., the state $\tilde{s}_i$ is the union of all states in $A_i$). We define the set of the states in the new discrete-time Markov chain as $Tr^a = \{\tilde{s}_i : l \le i \le l+d\}$. Note that the Markov chain is a substochastic process in which the process will exit $Tr^a$ eventually. Let the state into which the process exits be denoted by $\tilde{s}_{l-1}$.

Let $p_{ss'}$ be the transition probability from state $s$ to $s'$ in the original process $S(t)$. We define the transition probability from $\tilde{s}_i$ to $\tilde{s}_j$, $l \le i, j \le l+d$ in $Tr^a$ as

$$p_{\tilde{s}_i \tilde{s}_j} \triangleq \frac{\sum_{s \in A_i}\sum_{s' \in A_j} v_s p_{ss'}}{\sum_{s \in A_i} v_s} \quad i \ne j \quad (A10)$$

Furthermore, we define $p_{\tilde{s}_l \tilde{s}_{l-1}}$ as the probability the process transits to $\tilde{s}_{l-1}$ from $\tilde{s}_l$:

$$p_{\tilde{s}_l \tilde{s}_{l-1}} \triangleq \frac{\sum_{s \in A_l}\sum_{s' \in A_{l-1}} v_s p_{ss'}}{\sum_{s \in A_l} v_s} \quad (A11)$$

Write

$$\begin{aligned} p_{\tilde{s}_i \tilde{s}_i} &= 1 - p_{\tilde{s}_i \tilde{s}_{i-1}} - p_{\tilde{s}_i \tilde{s}_{i+1}} \quad l \le i < l+d \\ p_{\tilde{s}_{l+d} \tilde{s}_{l+d}} &= 1 - p_{\tilde{s}_{l+d} \tilde{s}_{l+d-1}} \end{aligned} \quad (A12)$$

Denote the one-step transition probability matrix composed of $p_{\tilde{s}_i \tilde{s}_j}$, $\forall \tilde{s}_i, \tilde{s}_j \in Tr^a$ by $\prod^a$.

Invoking (A10) and (A11), the transition probability from $\tilde{s}_{i+1}$ to $\tilde{s}_i$, $\forall l-1 \le i < l+d$, is

$$p_{\tilde{s}_{i+1} \tilde{s}_i} = \frac{\sum_{s \in A_{i+1}}\sum_{s' \in A_i} v_s p_{ss'}}{\sum_{s \in A_{i+1}} v_s} = \frac{\frac{(i+1)}{\upsilon_0}\sum_{s \in A_{i+1}} v_s}{\sum_{s \in A_{i+1}} v_s} = \frac{i+1}{\upsilon_0} \quad (A13)$$

Let $n_s$ denote the number of right neighbors of state $s$ in the trap $Tr$. Similarly, we have the transition probability from $\tilde{s}_i$ to $\tilde{s}_{i+1}$, $\forall l \leq i < l+d$, as

$$p_{\tilde{s}_i \tilde{s}_{i+1}} = \frac{\sum_{s \in A_i} \sum_{s' \in A_{i+1}} v_s p_{ss'}}{\sum_{s \in A_i} v_s} = \frac{\sum_{s \in A_i} v_s n_s \frac{\rho}{v_0}}{\sum_{s \in A_i} v_s}$$

Invoking Lemma 2, we get

$$p_{\tilde{s}_i \tilde{s}_{i+1}} = \frac{\sum_{s \in A_i} \left( \frac{c}{\rho^{l+d-|s|}} + o\left(\frac{c}{\rho^{l+d-|s|}}\right) \right) n_s \frac{\rho}{v_0}}{\sum_{s \in A_i} \left( \frac{c}{\rho^{l+d-|s|}} + o\left(\frac{c}{\rho^{l+d-|s|}}\right) \right)}$$

$$= \frac{\frac{c}{\rho^{l+d-i}} \sum_{s \in A_i} n_s \frac{\rho}{v_0} + \sum_{s \in A_i} n_s \frac{\rho}{v_0} o\left(\frac{1}{\rho^{l+d-i}}\right)}{\frac{c}{\rho^{l+d-i}} \sum_{s \in A_i} 1 + \sum_{s \in A_i} o\left(\frac{c}{\rho^{l+d-i}}\right)}$$

Noting that $\sum_{s \in A_i} 1 = |A_i|$ and $\sum_{s \in A_i} n_s = |A_{i+1}|(i+1)$, we have

$$p_{\tilde{s}_i \tilde{s}_{i+1}} = \frac{|A_{i+1}|}{|A_i|}(i+1)\frac{\rho}{v_0} + o\left(\frac{\rho}{v_0}\right). \quad (A14)$$

Combining (A13) and (A14), we can rewrite $\prod^a$ as

$$p_{\tilde{s}_i \tilde{s}_{i-1}} = \frac{i}{v_0}, \quad i = l, \cdots, l+d$$

$$p_{\tilde{s}_i \tilde{s}_{i+1}} = \frac{|A_{i+1}|}{|A_i|}(i+1)\frac{\rho}{v_0} + o\left(\frac{\rho}{v_0}\right) \quad i = l, l+1, \cdots, l+d-1 \quad (A15)$$

$$p_{\tilde{s}_i \tilde{s}_i} = 1 - p_{\tilde{s}_i \tilde{s}_{i-1}} - p_{\tilde{s}_i \tilde{s}_{i+1}} \quad i = l, \cdots, l+d$$

Note that $\prod^a$ is a substochastic matrix and it does not include $p_{\tilde{s}_l \tilde{s}_{l-1}} = \frac{l}{v_0}$.

2) To show that $\lambda_1$ is also an eigenvalue of matrix $\prod^a$

Define $v^a_{\tilde{s}_i} = \sum_{s \in A_i} v_s$, we next show that $\vec{V}_1^a = (v^a_{\tilde{s}_l}, v^a_{\tilde{s}_{l+1}}, \cdots, v^a_{\tilde{s}_{l+d}})$ is the eigenvector to $\prod^a$ with eigenvalue $\lambda_1$. That is, we show that

$$\vec{V}_1^a \prod^a = \lambda_1 \vec{V}_1^a \quad (A16)$$

i) $\forall l < i < l+d$

Starting from the LHS of (A16), we have

$$v^a_{\tilde{s}_i} p_{\tilde{s}_i \tilde{s}_i} + v^a_{\tilde{s}_{i+1}} p_{\tilde{s}_{i+1} \tilde{s}_i} + v^a_{\tilde{s}_{i-1}} p_{\tilde{s}_{i-1} \tilde{s}_i}$$

$$= \sum_{s \in A_i} v_s p_{ss} + \sum_{s \in A_{i+1}} \sum_{s' \in A_i} v_s p_{ss'} + \sum_{s \in A_{i-1}} \sum_{s' \in A_i} v_s p_{ss'}$$

$$= \sum_{s \in A_i} \left[ v_s p_{ss} + \sum_{s' \in A_{i+1}} v_{s'} p_{s's} + \sum_{s' \in A_{i-1}} v_{s'} p_{s's} \right] \quad (A17)$$

$$= \sum_{s \in A_i} \lambda_1 v_s = \lambda_1 v^a_{\tilde{s}_i}$$

The second line of (A17) is obtained from our definitions and (A10); the third line is obtained by exchanging the sequence of the sum operation; the fourth line is obtained by noting that in the original Markov process $S_{Tr}(t)$, we have $v_s p_{ss} + \sum_{s' \in A_{i+1}} v_{s'} p_{s's} + \sum_{s' \in A_{i-1}} v_{s'} p_{s's} = \lambda_1 v_s$.

ii) $i = l$ and $i = l+d$

Starting from the LHS of (A16), we have

$$v^a_{\tilde{s}_l} p_{\tilde{s}_l \tilde{s}_l} + v^a_{\tilde{s}_{l+1}} p_{\tilde{s}_{l+1} \tilde{s}_l}$$

$$= \sum_{s \in A_l} v_s p_{ss} + \sum_{s \in A_{l+1}} \sum_{s' \in A_l} v_s p_{ss'} = \sum_{s \in A_l} \left[ v_s p_{ss} + \sum_{s' \in A_{l+1}} v_{s'} p_{s's} \right] \quad (A18)$$

$$= \sum_{s \in A_l} \lambda_1 v_s = \lambda_1 v^a_{\tilde{s}_l}$$

Similarly, we can prove the case $i = l+d$.

Combining i) and ii), we have proved (A16). That is, $\lambda_1$ is an eigenvalue of matrix $\prod^a$ and $\vec{V}_1^a$ is the associated eigenvector.

3) Birth-death process $S_{Tr^a}(t)$

Next, we reverse the uniformization process. Based on the one-step transition matrix $\prod^a$ and $v_0$, we construct a continuous time Markov process $S_{Tr^a}(t)$ in which the transition rate between states are determined by

$$\mu(\tilde{s}_i \to \tilde{s}_{i-1}) = i \quad i = l+1, \cdots, l+d$$

$$\mu(\tilde{s}_i \to \tilde{s}_{i+1}) \triangleq c_i \rho = \frac{|A_{i+1}|}{|A_i|}(i+1)\rho + o(\rho) \quad i = l, l+1, \cdots, l+d-1 \quad (A19)$$

Furthermore, we have $\mu(\tilde{s}_l \to \tilde{s}_{l-1}) = l$.

The state-transition diagram of $S_{Tr^a}(t)$ is shown in Fig. A1.

In item 2), we have shown that $\lambda_1$ is an eigenvalue of matrix $\prod^a$ and $\vec{V}_1^a$ is the associated eigenvector. According to (A1), we obtain

$$T^a_{E|\vec{V}_1^a} = \frac{1}{v_0} \frac{1}{1-\lambda_1} \quad (A20)$$

where $T^a_{E|\vec{V}_1^a}$ is expected exit time of $S_{Tr^a}(t)$ given $P_0 = \vec{V}_1^a$.

It is easy to see that $S_{Tr^a}(t)$ is a Birth-Death process. Chapter 5 of [14] gives a closed-form expression for the expected passage time between states in a Birth-Death process, which provides another way to compute $T^a_{E|\vec{V}_1^a}$. Here we make use of this result.

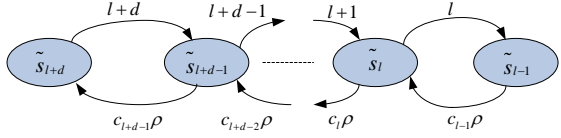

Fig.A1 The state transition process of $S_{Tr^a}(t)$

Similar to Eqn. (5.2.3) of [14], we define the potential coefficient of $\tilde{s}_{l+d}$ to be one. That is, $\theta_{l+d}=1$. The potentials of $\tilde{s}_i$, $\theta_i, i=l,\cdots,l+d-1$, are

$$\theta_i = \frac{\mu(\tilde{s}_{l+d}\to\tilde{s}_{l+d-1})\mu(\tilde{s}_{l+d-1}\to\tilde{s}_{l+d-2})\cdots\mu(\tilde{s}_{i+1}\to\tilde{s}_i)}{\mu(\tilde{s}_{l+d-1}\to\tilde{s}_{l+d})\mu(\tilde{s}_{l+d-2}\to\tilde{s}_{l+d-1})\cdots\mu(\tilde{s}_i\to\tilde{s}_{i+1})}$$

$$= \frac{(l+d)(l+d-1)\cdots(i+1)}{\left(\frac{|A_{l+d}|}{|A_{l+d-1}|}(l+d)\rho+o(\rho)\right)\left(\frac{|A_{l+d-1}|}{|A_{l+d-2}|}(l+d-1)\rho+o(\rho)\right)\cdots\left(\frac{|A_{i+1}|}{|A_i|}(i+1)\rho+o(\rho)\right)}$$

$$= \frac{(l+d)(l+d-1)\cdots(i+1)}{\rho^{l+d-i}\frac{|A_{l+d}|(l+d)(l+d-1)\cdots(i+1)}{|A_i|}+o(\rho^{l+d-i})} = \frac{|A_i|}{|A_{l+d}|}\rho^{i-l-d}+o(\rho^{i-l-d})$$

Invoking the equation above (5.2.4) in [14], we know the expected passage time from $\tilde{s}_l$ to $\tilde{s}_{l-1}$ can be computed as

$$T(\tilde{s}_l\to\tilde{s}_{l-1}) = \frac{1}{\mu(\tilde{s}_l\to\tilde{s}_{l-1})\theta_l}\sum_{k=l}^{l+d}\theta_k = \frac{1}{l\theta_l}\sum_{k=l}^{l+d}\theta_k$$

$$= \frac{1}{l\left(\frac{|A_l|}{|A_{l+d}|}\rho^{-d}+o(\rho^{-d})\right)}\left(1+\frac{|A_{l+d-1}|}{|A_{l+d}|}\rho^{-1}+\frac{|A_{l+d-2}|}{|A_{l+d}|}\rho^{-2}+\cdots+\frac{|A_l|}{|A_{l+d}|}\rho^{-d}+o(\rho^{-1})\right)$$

$$= \left(1+\frac{|A_{l+d-1}|}{|A_{l+d}|}\rho^{-1}+\frac{|A_{l+d-2}|}{|A_{l+d}|}\rho^{-2}+\cdots+\frac{|A_l|}{|A_{l+d}|}\rho^{-d}+o(\rho^{-1})\right)\left(\frac{|A_{l+d}|}{l|A_l|}\rho^d-\frac{o(\rho^d)}{l}\right)$$

$$= \frac{|A_{l+d}|}{l|A_l|}\rho^d+o(\rho^d)$$

Similarly, the expected passage times from $\tilde{s}_l$, $\tilde{s}_{l+1}$, $\cdots,\tilde{s}_{l+d}$ to $\tilde{s}_{l-1}$ satisfy

$$T(\tilde{s}_i\to\tilde{s}_{l-1}) = \frac{|A_{l+d}|}{l|A_l|}\rho^d+o(\rho^d), l\leq i\leq l+d. \quad(A21)$$

Equation (A21) indicates that the passage time from $\tilde{s}_l$, $\tilde{s}_{l+1}$, $\cdots,\tilde{s}_{l+d}$ to $\tilde{s}_{l-1}$ has the same highest-order term of $\rho$. Given arbitrary initial distribution including $\vec{V}_1^a$, the expected passage time should be $\frac{|A_{l+d}|}{l|A_l|}\rho^d+o(\rho^d)$. That is,

$$T^a_{E|\vec{V}_1^a} = \sum_{i=l}^{l+d}v_{\tilde{s}_i}T(\tilde{s}_i\to\tilde{s}_{l-1}) = \frac{|A_{l+d}|}{l|A_l|}\rho^d+o(\rho^d) \quad(A22)$$

Combining (A20) and (A22), we can write

$$T^a_{E|\vec{V}_1^a} = \frac{1}{1-\lambda_1}\frac{1}{\upsilon_0} = \frac{|A_{l+d}|}{l|A_l|}\rho^d+o(\rho^d).$$

That is, $\frac{1}{1-\lambda_1} = \frac{|A_{l+d}|}{l|A_l|}\rho^d\upsilon_0+o(\rho^d\upsilon_0)$. Back to our original Markov chain $S_{Tr}(t)$, we have $T_{E|\vec{V}_1} = \frac{|A_{l+d}|}{l|A_l|}\rho^d+o(\rho^d)$.

That is, $\beta = \frac{|A_{l+d}|}{l|A_l|}$.

**Step 5:** We want to show that $T(s\to B) = \beta\rho^d+o(\rho^d)$ $\forall s\in Tr$. That is, the highest-order term of $T(s\to B)$, $\beta\rho^d$, is the same as the highest-order term of $T_{E|\vec{V}_1}$.

We separate the proof into two steps: 1) prove that $T(s\to B) = \Theta(\rho^d)$ for all $s\in Tr$; 2) show that $t_s$ in $T(s\to B) = t_s\rho^d+o(\rho^d)$ are the same for all $s\in Tr$. Then, since $T_{E|\vec{V}_1}$ is of a convex combination of $T(s\to B)$ for different $s$, it is obvious that $t_s = \beta$.

We complete step 1) above by contradiction. The outline of our proof is as follows. Suppose that $T(s\to B)$ of different $s$ have different highest-order terms of $\rho$. We then construct a new trap based on this assumption and show that this new trap has contradicting properties. Lemmas 3-5 and definitions 1-3 below are introduced for the construction of the new trap $\bar{S}^*_{s^*}$ (the definition of which is given in Definition 4). Lemmas 6 and 7 are a set of contradicting results, and Lemma 8 completes the whole proof.

**Lemma 3:** In a trap $Tr$, there must be at least one state $s$ whose $T(s\to B) = \Omega(\rho^d)$.

**Proof**: From (A1) and (A6), we have
$$T_{E|\vec{V}_1} \triangleq \sum_{s_i\in Tr}v_{s_i}T(s_i\to B) = \beta\rho^d+o(\rho^d) \quad(A23)$$
where $\beta = \frac{|A_{l+d}|}{l|A_l|}$.

We prove the lemma by contradiction. Suppose that there is no state $s\in Tr$ satisfying $T(s\to B) = \Omega(\rho^d)$. That is, $\forall s\in Tr, T(s\to B) = o(\rho^d)$. Note that in (A23), $v_{s_i} = O(1)$, then $T_{E|\vec{V}_1} = \sum_{s_i\in Tr}v_{s_i}T(s_i\to B) = o(\rho^d)$. This contradicts (A23). □

For each state $s\in Tr$, let $T(s\to B) = \Theta(\rho^{x_s})$. We refer to $x_s$ as the exit-time exponent of state $s$. Let $x^*$ be the maximum exit-time exponent: $x^* = \max_{s\in Tr}(x_s)$. By Lemma 3, $x^*\geq d$.

Define $S^* = \{s \in Tr \mid T(s \to B) = \Theta(\rho^{x^*})\}$. That is, $S^*$ is the set of states in $Tr$ with the highest-order passage time.

**Definition 1:** Let $\bar{S}^* = Tr - S^*$. By definition, for each state $T(s \to B) = o(\rho^{x^*}) \forall s \in \bar{S}^*$.

**Definition 2:** For a state $s \in Tr$, define all left neighbors of state $s$ to be its children. We denote the children of state $s$ by $C_s$. Define all right neighbors of state $s$ to be its parents. We denote the parents of state $s$ by $P_s$. Define $m_s = |C_s| = |s|$ and $n_s = |P_s|$.

**Definition 3:** If there is a leftward path from state $s$ to state $s'$ (i.e., there is a sequence of leftward transitions leading from $s$ to $s'$ within the Markov chain), then we say that $s'$ is a descendant of $s$. We denote the set of descendants of state $s$ by $D_s$. If there is a rightward path from state $s$ to state $s'$ $s'$ (i.e., there is a sequence of leftward transitions leading from $s$ to $s'$ within the Markov chain), then we say that $s'$ is an ancestor of $s$.

**Lemma 4:** Consider any state $s^* \in S^*$. Then, $s \in S^* \ \forall \ s \in D_{s^*}$.

**Proof of the Lemma 4:** It suffices to prove $s \in S^* \ \forall \ s \in C_{s^*}$. Consider a state $s \in C_{s^*}$. The dynamic equation associated with state $s$ is

$$T(s \to B) = \frac{1}{n_s \rho + m_s} + \frac{1}{n_s \rho + m_s} \sum_{s' \in C_s} T(s' \to B)$$
$$+ \frac{\rho}{n_s \rho + m_s} \sum_{s' \in P_s} T(s' \to B) \quad (A24)$$
$$\geq \frac{\rho}{n_s \rho + m_s} T(s^* \to B)$$

Thus, $T(s \to B) = \Omega(\rho^{x^*})$ since $T(s^* \to B) = \Theta(\rho^{x^*})$. By definition, all states $s \in Tr$ must have $T(s \to B) = O(\rho^{x^*})$. Hence, $T(s \to B) = \Theta(\rho^{x^*}) \ \forall \ s \in C_{s^*}$ □

By our definition of traps, there is a path from any state $s \in S^*$ to any state $s' \in \bar{S}^*$ within the trap. This means that if $\bar{S}^* \neq \emptyset$, then we can find two states, $s \in S^*$ and $s' \in \bar{S}^*$, where $s$ and $s'$ are neighbors.

**Lemma 5:** If $\bar{S}^* \neq \emptyset$, then there exist $s \in S^*$ and $s' \in \bar{S}^*$, such that $s$ and $s'$ neighbors. Furthermore, for two such neighbors, $s$ must be a child of $s'$ and cannot be a parent of $s'$.

**Proof:** The first sentence is due to our definition of traps. The second sentence is a corollary of Lemma 4. □

**Definition 4:** In general, there could be one or more states $s \in \bar{S}^*$ with a child in $S^*$. These states could be located at different columns. We look at those at the rightmost columns (i.e., those with the largest cardinality $|s|$). We arbitrarily choose one of these states with the largest-cardinality. We call this state $s'$. We define a trap based on $s'$ as follows:

Consider all states in $\bar{S}^*$ that are connected to $s'$ via states in $\bar{S}^*$ only (i.e., we consider all $s \in \bar{S}^*$ such that there is a path $(s, s_1, s_2, ..., s_n, s')$, $s_i \in \bar{S}^*$, from $s$ to $s'$. Among these states, we discard those in a column to the left of $s'$. The remaining states, $\bar{S}^*_{s'} \subseteq \bar{S}^*$, forms a trap.

Denote the columns in $\bar{S}^*_{s'}$ by $A'_i$, where $A'_i \subseteq A_i$. Thus, $\bar{S}^*_{s'}$ is a trap with columns $A'_{|s'|}, A'_{|s'|+1}, ...$. In particular, note that all transitions out of $\bar{S}^*_{s'}$ must be through columns $A'_{|s'|}$. That is, exit from $\bar{S}^*_{s'}$ to outside of $\bar{S}^*_{s'}$ must be from a state in column $A'_{|s'|} \subseteq \bar{S}^*_{s'}$ to a state in column $A_{|s'|-1} \subseteq G$.

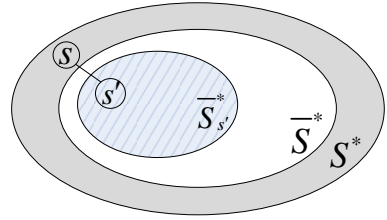

Fig. A2 Illustrating diagram showing the construction of $\bar{S}^*_{s'}$. We partition the state space of a trap into two sets: $S^*$ and $\bar{S}^*$. In $\bar{S}^*$ we form a new trap based on state $s'$ with the constraint that there is no path from any state in $\bar{S}^*_{s'}$ to states in $\bar{S}^* - \bar{S}^*_{s'}$. Thus, given the system is within $\bar{S}^*_{s'}$, in order to travel to the outside, it must first exit from $\bar{S}^*_{s'}$ from a state within the left column of $\bar{S}^*_{s'}$.

Let $\vec{V}' = (v'_{\tilde{s}'})_{\tilde{s}' \in \bar{S}^*_{s'}}$ be the normalized eigenvector associated with the trap $\bar{S}^*_{s'}$ associated with the largest eigenvalue, $\lambda'$.

Define the eigen-exit time to be:
$$T_{\vec{V}'B} = \sum_{\tilde{s}' \in \bar{S}^*_{s'}} v'_{\tilde{s}'} T(\tilde{s}' \to B) \quad (A25)$$

That is, $T_{\vec{V}'B}$ is the mean exit time to $B$ given the initial probability distribution $\vec{V}'$.

Next we prove $\bar{S}^* = \emptyset$ by contradiction. That is, $\forall s \in Tr, T(s \to B) = \Theta(\rho^{x^*})$.

**Lemma 6:** If $\bar{S}^* \neq \emptyset$, then $T_{\vec{V}'B} = o(\rho^{x^*})$.

**Proof:** If $\bar{S}^* \neq \emptyset$, then Definition 4 is valid. We can form the trap $\bar{S}^*_{s'}$. Note that in (A25), $v'_{\tilde{s}'} = O(1)$, and $T(\tilde{s}' \to B) = o(\rho^{x^*}) \; \forall \; \tilde{s}' \in \bar{S}^*_{s'}$ by definition. Thus, from (A25), we have $T_{\vec{V}'B} = o(\rho^{x^*})$ □

**Lemma 7:** If $\bar{S}^* \neq \varnothing$, then $T_{\vec{V}'B} = \Theta(\rho^{x^*})$.

**Proof:** We can write $T_{\vec{V}'B}$ as summation of two terms:
$$T_{\vec{V}'B} = T_{\vec{V}'E} + T_{EB} \quad (A26)$$
where $T_{\vec{V}'E}$ is the expected time the system stays within $\bar{S}^*_{s'}$ before the first exit from $\bar{S}^*_{s'}$ to $A_{|s'|-1}$, given the initial distribution $\vec{V}'$; $T_{EB}$ is the remaining expected time after the first exit from $\bar{S}^*_{s'}$ to $A_{|s'|-1}$ until the system reaches $B$. Fig. A3 shows an illustrating diagram for $T_{\vec{V}'E}$ and $T_{EB}$.

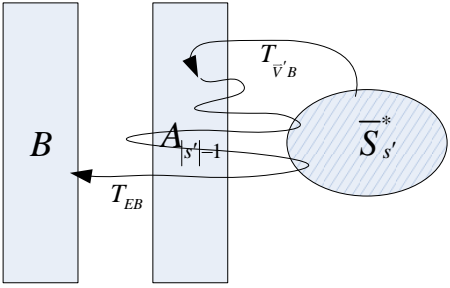

Fig. A3 Illustrating diagram for $T_{\vec{V}'E}$ and $T_{EB}$. Note that in general, after the first exit, it is possible for the system to evolve back and forth into and out of the trap $\bar{S}^*_{s'}$ several times before reaching $B$ eventually.

Let $P_{\tilde{s}|\vec{V}'}$ be the probability that the first exit from $\bar{S}^*_{s'}$ is to state $\tilde{s} \in A_{|s'|-1}$. We can write
$$T_{EB} = \sum_{\tilde{s} \in A_{|s'|-1}} P_{\tilde{s}|\vec{V}'} T(\tilde{s} \to B) \geq P_{s|\vec{V}'} T(s \to B) \quad (A27)$$
where $s \in S^*$ is a child of $s'$ in $S^*$.

If we can show that $P_{s|\vec{V}'} = \Theta(1)$. Then, (A26), (A27), and the fact $T(s \to B) = \Theta(\rho^{x^*})$ immediately gives $T_{\vec{V}'B} = \Theta(\rho^{x^*})$ and we are done with the proof.

To prove $P_{s|\vec{V}'} = \Theta(1)$, we first note that $P_{s|\vec{V}'} \geq P_{s' \to s|\vec{V}'}$, where $s' \to s$ is the event that the exit pattern is $s' \to s$ (i.e., $P_{s' \to s|\vec{V}'}$ is the probability that the first exit is via $s'$ within the trap to $s$ outside the trap, given the initial distribution $\vec{V}'$). In the next paragraph, we prove that $P_{s' \to s|\vec{V}'} = \Theta(1)$. Then, since $P_{s|\vec{V}'} = O(1)$ (by definition of probability), we have $P_{s|\vec{V}'} = \Theta(1)$.

Let $v_{s'}(t)$ be the probability that the system is still in state $\tilde{s}'$ at time $t$ given the initial distribution $\vec{V}'$. If we prepare a large ensemble of experiments, say $N$ of them, with initial condition $\vec{V}'$, and observe their exit patterns. Among all exits, the number of exit from state $s'$ to state $s$ is $N \int_0^{\infty} v_{s'}(t) \cdot 1 \, dt$ (recall that the transition rate for $s' \to s$ is 1). The total number of exits from all states in column $A'_{|s'|}$ is $N \sum_{\tilde{s}' \in A'_{|s'|}} \int_0^{\infty} v_{\tilde{s}'}(t) |s'| \, dt$. Thus, the probability that in any of the experiment, the exit is of the pattern $s' \to s$ is
$$P_{s' \to s|\vec{V}'} = \frac{N \int_0^{\infty} v_{s'}(t) dt}{N \sum_{\tilde{s}' \in A'_{|s'|}} \int_0^{\infty} v_{\tilde{s}'}(t) |s'| \, dt} = \frac{\int_0^{\infty} v_{s'}(t) dt}{\sum_{\tilde{s}' \in A'_{|s'|}} \int_0^{\infty} v_{\tilde{s}'}(t) |s'| \, dt} \quad (A28)$$

With initial condition $\vec{V}'$, $v_{\tilde{s}'}(t) = v_{\tilde{s}'}$ when $t = 0$. In the uniformed discrete-time Markov chain, after $n$ transition, the probability that the system is still in state $\tilde{s}'$ is $v_{\tilde{s}'} \lambda'^n$ (i.e., $v_{\tilde{s}'}[n] = v_{\tilde{s}'} \lambda'^n$) according to the definition of $\vec{V}'$. According to the uniformization technique introduced in Step 1, the epochs at which transitions occur are a Poisson process with rate $\upsilon_0$. Then, the probability of having $n$ transitions in time $t$ is $P_n(t) = \frac{(\upsilon_0 t)^n}{n!} e^{-\upsilon_0 t}$. Thus, we have
$$v_{s'}(t) = \sum_n v_{s'}[n] P_n(t) = \sum_n v_{s'} \lambda'^n \frac{(\upsilon_0 t)^n}{n!} e^{-\upsilon_0 t}$$
$$= v_{s'} e^{-\upsilon_0 t} \cdot \sum_n \lambda'^n \frac{(\upsilon_0 t)^n}{n!} = v_{s'} e^{-\upsilon_0 t} \cdot e^{\lambda' \upsilon_0 t} \sum_n \frac{(\lambda' \upsilon_0 t)^n}{n!} e^{-\lambda' \upsilon_0 t}$$
$$= v_{s'} e^{-(1-\lambda')\upsilon_0 t}$$
and $\int_0^{\infty} v_{\tilde{s}'}(t) dt = \frac{v_{\tilde{s}'}}{(1-\lambda')\upsilon_o}$, $\forall \; \tilde{s}' \in A'_{|s'|}$.

Substituting into (A28), we have
$$P_{s' \to s|\vec{V}'} = \frac{v_{s'}}{|s'| \sum_{\tilde{s}' \in A'_{|s'|}} v_{\tilde{s}'}} \quad (A29)$$

Invoking Lemma 2, we have $P_{s' \to s|\vec{V}'} = \frac{1}{|s'||A'_{|s'|}|} + o(1)$. That is, $P_{s' \to s|\vec{V}'} = \Theta(1)$. □

**Lemma 8:** $T(s \to B) = \Theta(\rho^d), \forall s \in Tr$

**Proof:** A corollary of Lemmas 6 and 7 is $T(s \to B) = \Theta(\rho^{x^*}), \forall s \in Tr$ because $\bar{S}^* = \varnothing$. From (A1) and (A6), we can immediately deduce that $x^*$ cannot be larger than $d$ or smaller than $d$. Thus, $x^* = d$. □

Next, we prove that the $t_s$ in $T(s \to B) = t_s \rho^d + o(\rho^d)$ are the same for all $s \in Tr$.

For each $s \in Tr$, we can write

$$T(s \to B) = \frac{1}{m_s + n_s \rho} + \sum_{s' \in C_s} \frac{1}{m_s + n_s \rho} T(s' \to B) + \sum_{s' \in P_s} \frac{\rho}{m_s + n_s \rho} T(s' \to B) \quad (A30)$$

i) If $s$ has no right neighbors (i.e., $n_s = 0$ and $P(s) = \varnothing$), then matching the highest-order term in (A30), we have

$$t_s = \frac{1}{m_s} \sum_{s' \in C_s} t_{s'} \quad (A31)$$

ii) If $n_s \neq 0$, then matching the highest-order term in (A30) gives

$$t_s = \frac{1}{n_s} \sum_{s' \in D_s} t_{s'} \quad (A32)$$

Eqn (A31) and (A32) can be written into a matrix form

$$T = PT \quad (A33)$$

where $T = (t_s)_{s \in Tr}$ is a $|Tr| \times 1$ column vector, and $P$ is a stochastic matrix. From (A33), we can form a substochastic process as follows. We choose one of the states in $Tr$ as the state in $U$ and the other states are in $V$. Let the chosen state be the state corresponding to the bottom element in $T$. Denote it by $s*$.

We remove the last row on the LHS and the RHS of (A33). We end up with a dynamic equation as follows:

$$T' = P'T' + E' \quad (A34)$$

where $T'$ is a $(|Tr| - 1) \times 1$ column vector consisting of $T$ with the bottom element removed; $P'$ is a $(|Tr| - 1) \times (|Tr| - 1)$ substochastic matrix which is $P$ with the bottom row and the rightmost column removed; and $E'$ is a $(|Tr| - 1) \times 1$ column vector whose elements are in general function of $t_{s'}$.

If we replace the dynamic equation (A34) by

$$T' = P'T' + \begin{pmatrix} 1 \\ \vdots \\ 1 \end{pmatrix} \quad (A35)$$

we end up having a dynamic equation governing the mean exit time of a discrete-time Markov chain from states in $V$ to the state in $U$. The Markov chain is such that all states in $V$ has a path to $U$. Thus, eventually, regardless of which state the process starts out with in $V$, it will exit to $U$. The $t_s$ solved using (A35) is the mean exit time with the initial state being $s \in V$.

In particular,

$$T' = (I - P')^{-1} \begin{pmatrix} 1 \\ \vdots \\ 1 \end{pmatrix} = (I + P' + P'^2 + ...) \begin{pmatrix} 1 \\ \vdots \\ 1 \end{pmatrix}$$

and that $(I - P')^{-1}$ is well defined and exist (Write $A = I - P'$ and then we have $I - A = P'$. Note that $P'$ is a substochastic matrix for a connected Markov chain. Thus, $\lim_{i \to \infty}(I - A)^i = \lim_{i \to \infty} P'^i = 0$. According to the Theorem of Matrix Inversion by Neumann Series [23], we have $A$ is invertible and $A^{-1} = (I - P')^{-1} = \sum_{i=0}^{\infty} P'^i$). This means that in the original eqn (A34), we can write

$$T' = (1 - P')^{-1} E \quad (A36)$$

In other words, $t_s$ as a function of $t_{s*}$ is governed by eqn (A36). The solution exists and is unique. Thus, if we can guess a solution of $t_s$ as a function $t_{s*}$ that satisfies (A36), it must be the only solution. Eqn (A36) comes from (A31) and (A32) originally. We guess $t_s = t_{s*}$ and this solution satisfies (A31) an (A32). Q.E.D.

## APPENDIX B: PROOF OF THE THEOREM 2

*Proof of Theorem 2:* Consider a particular trap in a general network over a very long time $T$. Ref. [5] proved that the stationary probability distribution of the system is insensitive to the distributions of the backoff time and the transmission time given their means, so the stationary probability for the process to be within the trap is also insensitivity to the distributions. We can express

$$\Pr\{Tr\} = \sum_{s \in Tr} P_s$$

Suppose that over the time horizon, there are $n_e(T)$ times that the process enters the trap $Tr$. Since $T_V(Tr)$ is the ergodic sojourn time of $Tr$, we can write

$$\Pr\{Tr\} = \frac{T_V(Tr) \times n_e(T)}{T} \quad (B1)$$

To show that $T_V(Tr)$ is insensitive to the distributions of the backoff time and the transmission time, it suffices to show that $n_e(T)$ is insensitive to the distributions.

Note that the process can only visit the trap through the states in column $l-1$, then $n_e(T)$ can be expressed as

$$n_e(T) = n_{A_{l-1} \to A_l}(T) = \sum_{s \in A_{l-1}} \sum_{s' \in A_l} n_{ss'}(T) \tag{B2}$$

where $n_{ab}(T)$ denotes the number of transitions of type $a \to b$ observed over the horizon $T$.

Let $G(tr)$ denote the cumulative distribution of the transmission time (i.e., $G(tr) = \int_0^{tr} g(t_{tr}) dt_{tr}$ where $g(t_{tr})$ is the probability density function of $t_{tr}$). Let link $j$ be the link that is transmitting in state $s'$ while idle in state $s$. Then the transmission time of link $j$ is distributed as $G(rt)$ upon the system first entering $s'$ from $s$. We give the following lemma:

*Lemma 9:* For any state $s'$ in the left-most column of $Tr$, if $s$ is one of its left neighbors, we have

$$n_{ss'}(T) = T(s')/E[t_{tr}] = P_s T / E[t_{tr}] \tag{B3}$$

in which $T(s')$ is the amount of time within $[0,T]$ during which the system is in state $s'$.

*Proof:*
Consider a particular link and only the sub-time intervals within $[0,T]$ during which it is active. It is well known from renewal theory that the residual transmission time is $g_{RT}(rt) = \frac{1}{E[T_{tr}]}(1 - G(rt))$.

Note that a corollary of Theorem B1 in our previous technical report [5] is the invariant residual-time distributions property: the remaining backoff and transmission times of different links are independent, and therefore the fact that a link is counting down to zero or completing its transmission, thus experiencing a transition, has no bearing on the residual countdown and transmission times of other links.

If in state $s'$, link $j$ completes its transmission before other state transitions, then the next transition will be back to state $s$. If not, suppose that the system moves from state $s'$ to state $s'' \neq s$ because of the state transition of another link $k \neq j, i$. We know that the probability that the residual transmission time of link $j$ at the transition instant of link $k$ is between $x$ and $x + dx$ is $g_{RT}(x)dx$. The number of such transitions observed over $T$ is $n_{s's''}(T) g_{RT}(x) dx$.

By time reversibility (proved in [5]), $n_{s's''}(T) = n_{s''s'}(T)$. By the invariant residual-time property, the number of transitions of type $s'' \to s'$ and with residual time between $x$ and $x + dx$ is $n_{s's''}(T) g_{RT}(x) dx = n_{s''s'}(T) g_{RT}(x) dx$. Thus, for each transition $s' \to s''$, there is a corresponding transition $s'' \to s'$ with the same residual transmission time. It is as if the transmission continues where it left off. Thus, the number of times the transmission of link $j$ is completed while in state $s'$ (hence causing the transition $s' \to s$) during $T$ is $T(s')/E[t_{tr}] = P_s T / E[t_{tr}]$.

Plugging (B3) into (B2), it is then obvious that $n_e(T)$ is insensitive to the distributions of the backoff time and the transmission time given their means. □

## APPENDIX C: APPROXIMATE COMPUTATION OF ERGODIC SOJOURN TIME OF $T_V(Tr)$

Below shows that the state aggregation technique used to compute the ergodic sojourn time of a trap in Section V-D is approximate and not exact. However, it is exact when the states in the same column of the trap have the same number of right neighbors.

Consider a state $s_i$ with the trap $Tr$. Let $R(s_i)$ and $L(s_i)$ be the sets of its right neighbors and left neighbors, respectively. We have $R(s_i) = n_{s_i}$ and $L(s_i) = |s_i|$, where $n_{s_i}$ is the number of right neighbors of $s_i$. The evolution of the probability that the system is in state $s_i$, $P_{s_i}(t)$, is given by the differential equation:

$$\frac{dP_{s_i}(t)}{dt} = -(|s_i| + n_{s_i}\rho) P_{s_i}(t) + \sum_{s \in L(s_i)} \rho P_s(t) + \sum_{s \in R(s_i)} P_s(t) \tag{C1}$$

where $-(|s_i| + n_{s_i}\rho) P_{s_i}(t)$ is the contribution due to the events that the process leaves $s_i$ either to one of its $|s_i|$ neighbors at rate 1 or to one of its $n_{s_i}$ right neighbors at rate $\rho$; $\sum_{s \in L(s_i)} \rho P_s(t)$ is the contribution due to the transitions from its left neighbors to $s_i$ (at rate $\rho$) and $\sum_{s \in R(s_i)} P_s(t)$ is the contribution due to the transitions from its right neighbors to $s_i$ (at rate 1).

As can be seen in (C1), even if $P_s(t)$ for all $s$ in the same column of the trap are equal, this property may not be preserved as time evolves (unless $P_s(t)$ is already the stationary probability).

To see this, suppose that $P_s(t)$ for all $s$ in the same column of the trap are equal. The above can be written as

$$\frac{dP_{s_i}(t)}{dt} = -(|s_i| + n_{s_i}\rho) P_{s_i}(t) + |s_i| \rho P_{s_{i-1}}(t) + n_{s_i} P_{s_{i+1}}(t)$$

Note that the term $-n_{s_i}\rho P_{s_i}(t) + n_{s_i} P_{s_{i+1}}(t)$ is dependent on $n_{s_i}$, which may not be the same for all states $s_i$ within the same column. Thus, $\frac{dP_{s_i}(t)}{dt}$ are not the same for the states within the same column even if $P_{s_{i-1}}(t)$ and $P_{s_{i+1}}(t)$ are.

If the Markov chain is "uniform" in the sense that $n_{s_i}$ are the same for all states $s_i$ in the same column, then the above expression of $\frac{dP_{s_i}(t)}{dt}$ is the same for all the states in the same column given that $P_s(t)$ for all $s$ in the same column of the trap are equal. From (9), this condition applies when $t=0$. Thus, for a "uniform" Markov chain, the property is preserved for all time $t \geq 0$. The state aggregation technique becomes exact rather than approximate. Then we have the following lemma:

***Lemma 10:*** Given the initial condition is specified by (9), the states in the same column of $Tr$ have equal probability during transience if $n_{s_i}$ are the same for all states $s_i$ in the same column of the trap. Then the computation of (14) is exact.

## APPENDIX D: FIRST PASSAGE TIME FROM A TRAP TO ANOTHER TRAP

This appendix considers the first passage time from a trap to another trap, which gives further information on the duration of temporal starvation in some networks.

*1) Definition of the expected first passage time between traps:*

Let $l^*$ denote the left-most column of the trap $Tr_i$ (note that the columns of the trap are indexed with respect to the overall state-transition diagram), according to our definition of traps, when the process first visits $Tr_i$, the state it arrives at must be in column $l^*$. Assuming the system process is ergodic, we would like to derive the probability of a visit to $Tr_i$ beginning at state $s \in Tr_i$. Define $B_i \triangleq G \setminus Tr_i$. Let $h_{B_i,s}$ be the average number of visits to $Tr_i$ per unit time that begins at state $s \in Tr_i$, defined as follows:

$$h_{B_i,s} = \lim_{t \to \infty} \frac{1}{t}\left[\text{the number of transitions from } B_i \text{ to } s \text{ in } (0,t)\right]$$
$$= \sum_{s' \in B_i} P_{s'} \upsilon_{s's}$$
(D1)

where $\upsilon_{s's}$ is the transition rate from state $s'$ to state $s$ in the complete continuous-time Markov chain.

Given the fact that the system just arrives at the trap, we specify the initial distribution as

$$P_s(0) = \frac{h_{B_i,s}}{\sum_{s' \in Tr_i} h_{B_i,s'}}, \quad s \in Tr_i$$
$$P_s(0) = 0, \quad s \in B_i$$
(D2)

The first passage time from $Tr_i$ to $Tr_j$ is defined as the time for the system to arrive at $Tr_j$ given that the initial condition is specified by (D2):

$$T_P(Tr_i \to Tr_j) = \sum_{s \in Tr_i} P_s(0) T(s \to Tr_j) \quad (D3)$$

Similar to the derivation after the definition of the ergodic sojourn time of a trap in Section V-B, (D3) can be written as

$$T_P(Tr_i \to Tr_j) = \sum_{s \in Tr_i, |s|=l^*} \frac{1}{|A_{l^*}|} T_{sTr_j} \quad (D4)$$

*2) Approximate computation of $T_P(Tr_i \to Tr_j)$*

The computation of $T_P(Tr_i \to Tr_j)$ is more tricky than the computation of the ergodic sojourn time of a trap. In general, for two traps $Tr_i$ and $Tr_j$, we have

$$T_P(Tr_i \to Tr_j) \geq T_V(Tr_i) \quad (D5)$$

This is because after the process exits $Tr_i$, it may not transit to the trap $Tr_j$ immediately (indeed, it may revisit $Tr_i$ without traversing $Tr_j$ at all).

Based on the analysis of trap duration conducted in Section V, we propose an approximate method to compute the expectation of the first passage time from a trap to another trap, $T_P(Tr_i \to Tr_j)$. Note that if one trap is a subset of another trap, we define that the expected passage time between them is zero.

Consider two traps $Tr_i$ and $Tr_j$. In the complete state-transition diagram we find all the traps that have no intersections with $Tr_i$ or $Tr_j$ while making sure that each state is included in at most one trap. Denote the set of these traps in conjunction with $Tr_i$ and $Tr_j$ by $\mathbf{Tr} = \{Tr_1, Tr_2, \cdots, Tr_n\}$.

Next we construct a simplified stochastic process $\hat{S}(t)$ to compute the passage time between traps. We aggregate all the states in each trap $Tr \in \mathbf{Tr}$ into a single state, denoted by $s_{Tr_i}$, $i=1,\cdots,n$ and transform the complete state-transition diagram $G$ to a simplified state-transition diagram $G^*$.

Refer to our example in Fig. 1(a). The state transition diagram is shown in Fig.2. There are two traps in $G^{(1)}$, $Tr_1 \triangleq G_1^{(2)} = Tr\{\{1,4,5\},2\}$ and $Tr_2 \triangleq G_2^{(2)} = Tr\{\{2,3,6\},2\}$. The two columns are connected through column 1. By aggregating the states of a trap into a single state, we have the simplified state transition diagram shown in Fig. D1.

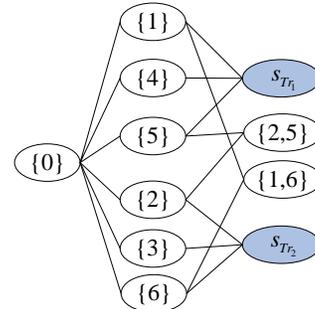

Fig.D1 Simplified State transition diagram of the network shown in Fig.1(a)

Let $\upsilon_s$ be the rate with which $S(t)$ makes a transition when in state $s$ in the simplified process. For a non-trap state $s$, we define $\upsilon_s$ as

$$\hat{\upsilon}_s = \upsilon_s \tag{D6}$$

where $\upsilon_s$ is the transition rate from state $s$ in the original state-transition diagram.

The transition rate from a non-trap state to a trap state $s_{Tr_i}$ in the simplified process is defined as

$$\hat{\upsilon}_{ss_{Tr_i}} = \sum_{s' \in Tr_i} \upsilon_{ss'} \tag{D7}$$

The transition rate from a non-trap state to another non-trap state is defined as

$$\hat{\upsilon}_{ss'} = \upsilon_{ss'} \tag{D8}$$

For trap state $s_{Tr_i}$, we define the rate the process $S(t)$ leaves the state as

$$\hat{\upsilon}_{s_{Tr_i}} = 1/T_V(Tr_i) \tag{D9}$$

where $T_V(Tr_i)$ can be approximately computed by (14).

Recall that the ergodic sojourn time of a trap $T_V(Tr_i)$ and the expected first passage time from $Tr_i$ to $Tr_j$, $T_P(Tr_i \to Tr_j)$ starts with the same initial condition specified by (10), this validates our treatment of (D9).

Assuming that the sojourn times of the traps are exponentially distributed, $\hat{S}(t)$ is therefore a continuous-time Markov process. Let $N(s)$ be the set of neighbors of state $s$ in $G^*$. The state transition probabilities upon a state transition can be specified as follows:

i) For a non-trap state $s$, $\forall s', s_{Tr_i} \in N(s)$, we have

$$\begin{cases} p_{ss'} = \hat{\upsilon}_{ss'} / \hat{\upsilon}_s \\ p_{ss_{Tr_i}} = \hat{\upsilon}_{ss_{Tr_i}} / \hat{\upsilon}_s = \sum_{s' \in Tr_i} \upsilon_{ss'} / \hat{\upsilon}_s \end{cases} \tag{D10}$$

Take state $\{1\}$ in Fig. D1 as an example. It has three neighbors, $s_{Tr_1}$, states $\{0\}$ and $\{1,6\}$. The state transition probability can be computed as

$$\begin{cases} p_{\{1\}\{0\}} = \hat{\upsilon}_{\{1\}\{0\}} / \hat{\upsilon}_{\{1\}} = 1/(1+3\rho) \\ p_{\{1\}s_{Tr_1}} = \hat{\upsilon}_{\{1\}s_{Tr_1}} / \hat{\upsilon}_{\{1\}} = 2\rho/(1+3\rho) \\ p_{\{1\}\{1,6\}} = \hat{\upsilon}_{\{1\}\{1,6\}} / \hat{\upsilon}_{\{1\}} = \rho/(1+3\rho) \end{cases} \tag{D11}$$

ii) For a trap state $s_{Tr_i}$:

As shown in (11), when the system process enters a trap $Tr$ with the left-most column is $l^*$, it begins at each state in column $l^*$ of the trap with equal probability. By time reversibility of $S(t)$ [5], when the process exits the trap, it exits from each state in column $l^*$ with equal probability, too. For a particular state $s$ in column $l^*$, it can exit through $l^*$ possible transitions to states in column $l^* - 1$. The transitions have equal probability. Thus, for a particular state $s'$ in column $l^* - 1$ that is connected to the trap, we have

$$p_{s_{Tr}s'} = \sum_{s \in A_{l^*}} P(\text{exits trap via } s)p_{ss'}$$
$$= \sum_{s \in A_{l^*}} p_{ss'} / |A_{l^*}| = n_{s'Tr} / |A_{l^*}| l^* \tag{D12}$$

where $n_{s'Tr}$ is the number of possible transitions through which state $s'$ can enter the trap $Tr$.

Refer to our example in Fig. 1(a), it turns out that $|A_{l^*}| = 3$, $l^* = 2$ and $n_{\{1\}Tr_1} = 2$. That is, in our example we have

$$p_{s_{Tr_1}\{1\}} = p_{s_{Tr_1}\{4\}} = p_{s_{Tr_1}\{5\}} = 1/3 \tag{D13}$$

After specifying the transition rates of states and state transition probabilities for $S(t)$, we can now compute the expected passage time from $s_{Tr_i}$ to $s_{Tr_j}$. For each state $s$ in $G^*$, define $e_s$ = E[waiting time until system enters $s_{Tr_j}$ |current state is $s$]. Conditioned on the first jump of the process, for each state $s$ we can write

$$\begin{cases} e_s = T_V(s) + \sum_{s' \in N(s)} p_{ss'} e_{s'}, & s \neq s_{Tr_j} \\ e_{s_{Tr_j}} = 0 \end{cases} \tag{D14}$$

where

$$T_V(s) = 1/\hat{\upsilon}_s. \tag{D15}$$

After solving the linear equations in (D10) for all states $s$, we obtain $e_{s_{Tr_i}}$, which is the expected passage time from $s_{Tr_i}$ to $s_{Tr_j}$ in the simplified state-transition diagram. The first passage time from $Tr_i$ to $Tr_j$, $T_P(Tr_i \to Tr_j)$, is then approximated by $e_{s_{Tr_i}}$.

*3) Simulation validation of our approximate computation*

Refer to our example shown in Fig. 1(a), there are two island-states $Tr_1 = Tr(\{1,4,5\},2)$ and $Tr_2 = Tr(\{2,3,6\},2)$. In simulations we collect the expected first passage time from $Tr_1$ to $Tr_2$, $T_P(Tr_1 \to Tr_2)$. By (D9) we can compute the expected first passage time from $Tr_1$ to $Tr_2$ approximately. The simulated $T_P(Tr_1 \to Tr_2)$ and computed $T_P(Tr_1 \to Tr_2)$ in terms of timeslots with respect to $\rho$ are compared in Table III. As can be seen, our approximate computation matches well with the simulated values.

Table III $T_P(Tr_1 \to Tr_2)$ with respect to $\rho$ in the network of Fig. 1(a)

| $\rho$ | $\rho_0$ | $5\rho_0$ | $10\rho_0$ | $100\rho_0$ |
|---|---|---|---|---|
| Simulated $T_P$ (ms) | 12.18 | 45.14 | 84.59 | 809.5 |
| Computed $T_P$ (ms) | 12.30 | 44.42 | 84.57 | 807.5 |

Furthermore, we consider larger 2-D grid networks. The simulated first passage time and computed passage time between the two traps are compared in Table IV. As can be seen, when $\rho$ is large (e.g., $\rho = 10\rho_0$), the accuracy of our computation is acceptable for all the networks tested. Besides that, Table IV shows that the expected first passage time of traps in the 2-D grid networks increases quickly with the network size, indicating more severe temporal starvation in the network.

Table IV $T_P(Tr_1 \to Tr_2)$ with respect to network size of the $m \times n$ 2-D grid network ($\rho = 10\rho_0$)

| $m \times n$ | $2\times 3$ | $2\times 4$ | $2\times 5$ | $3\times 4$ | $4\times 4$ |
|---|---|---|---|---|---|
| Simulated $T_P$ | 84.59 | 140.77 | 187.28 | 263.58 | 2035.7 |
| Computed $T_P$ | 84.57 | 135.90 | 177.72 | 266.26 | 1966.1 |

Finally, we reused the ten 20-links random networks generated in Section V-F to examine the accuracy of our approximation above. In each simulation run we gather the statistics of $T_P(Tr_1 \to Tr_2)$ and compare them with our computation using the method proposed in item 2).

Define $\Delta T_P$ as the ratio between prediction error and the simulated first passage time from a trap to another. Averaging over ten networks, we find that our approximate computation above can achieve roughly 90.43% of accuracy with $\rho = 10\rho_0$.

## APPENDIX E: UNIFIED DEFINITION OF STARVATION AND SUFFICIENT CONDITIONS FOR STARVATION IDENTIFICATION

This section proposes a unified definition of starvation, which attempts to capture both equilibrium and temporal starvations. We provide a sufficient condition in terms of a set of special traps to identify starvation with respect to this unified definition.

### 1) Unified definition of starvation

Let us consider the intervals between successive transmissions by link $i$. Specifically, let $T_i^{(j)}$ denote the instance at which link $i$ begins to transmit its $j^{th}$ packet. For convenience, let $T_i^{(0)} = 0$. Let $Y_i^{(j)} \triangleq T_i^{(j)} - T_i^{(j-1)}$ be the random variable representing the $j^{th}$ interval. For stationary process, we can drop the superscript. Let the probability distribution of $Y_i$ be denoted by $F_{Y_i}(y_i) \triangleq \Pr[Y_i \leq y_i]$. Recall that when a link is counting down during its backoff stage, the countdown process may be frozen when it senses a neighbor link transmitting. The interval $Y_i$ consists of the sum of one packet transmission time, the active countdown time, and the frozen time.

A possible notion for starvation is as follows. Suppose one chooses a random point in time to observe link $i$, and see how long it takes before link $i$ gets to transmit its next packet. If this time is excessive, we then said that link $i$ is starved. This time corresponds to the residual time of the interval $Y_i$, denoted by $X_i$. The probability density of $X_i$ is given by $f_{X_i}(x_i) = (1 - F_{Y_i}(x_i))/E[Y_i]$. We could define starvation as follows: link $i$ is said to be prone to starvation if $\Pr[X_i \leq x_i^{\text{target}}] < \varepsilon_i^{\text{target}}$ for some positive $x_i^{\text{target}}$ and $\varepsilon_i^{\text{target}}$. In this paper, for simplicity, we opt for an alternative definition instead, as follows:

*Unified Definition of starvation:* A link is starved if its residual countdown time $X_i$ satisfies

$$E[X_i] > \overline{X}_{\text{target}} \qquad (E1)$$

for some target mean $\overline{X}_{\text{target}} > 0$ determined by the requirements of the application running on top of the wireless network.

Note that $E[X_i] = \int_0^\infty x_i(1 - F_{Y_i}(x_i)dx_i / E[Y_i]$. By integration by part, we get

$$E[X_i] = \frac{1}{2E[Y_i]} \left( x_i^2(1 - F_{Y_i}(x_i)) \Big|_0^\infty + \int_0^\infty x_i^2 f_{Y_i}(x_i)dx_i \right) = \frac{E[Y_i^2]}{2E[Y_i]}.$$

Thus, as an alternative, we could also look at the second moment of the interval $Y_i$ to see if it is excessive compared with $E[Y_i]$. Indeed, ref. [20] uses this alternative definition.

The reason why starvation should be related to the second moment of $Y_i$ can be seen intuitively as follows. For a given $E[Y_i]$, the throughput of link $i$ is $1/E[Y_i]$. If the second moment $E[Y_i^2]$ is small, then link $i$ receives its service in a regular manner. On the other hand, if $E[Y_i^2]$ is large, it is more likely for our observation point to fall into a large interval, and as a result we need to wait a long time before the link transmits its next packet. This notion of starvation basically relates "suffering from starvation" to "receiving highly fluctuating service over time".

### 2) Procedure to identify frozen traps for a particular link

If link $i$ receives zero throughput within a trap $Tr$, we say $Tr$ is a frozen trap of link $i$. The procedure to identify frozen traps is as follows: first, all the traps in the network are identified using the procedure described in Section IV-A. We then go through all the traps and find out those in which link $i$ receives zero throughputs.

It is obvious that a frozen trap $T_f(i)$ of link $i$ does not contain a state in which link $i$ is transmitting. That it also does not contain a state in which link $i$ is actively counting down can be seen as follows. Suppose that $T_f(i)$ includes a state $s$ in which link $i$ is actively counting down. By definition, $T_f(i)$ includes all the right states of state $s$, among which there must be a state in which link $i$ is transmitting. But this contradicts with the fact that $T_f(i)$ does not contain such as a state. Thus, link $i$ is frozen in all states in $T_f(i)$.

Let $\mathcal{T}_f(i)$ be the set of frozen traps with respect to $i$:

$$\mathcal{T}_f(i) = \{T_f(i) | Th(i|T_f(i)) = 0\}$$

We consider the computation of $E[X_i]$:

$$\begin{aligned}
E[X_i] &= \sum_s p(s) E[X_i | S = s] \\
&= \sum_{T_f(i) \in \mathcal{T}_f(i)} \sum_{s \in T_f(i)} p(s) E[X_i | S = s] \\
&\quad + \sum_{s \notin T_f(i) \forall T_f(i) \in \mathcal{T}_f(i)} p(s) E[X_i | S = s] \\
&\geq \sum_{T_f(i) \in \mathcal{T}_f(i)} \sum_{s \in T_f(i)} p(s) E[X_i | S = s] \\
&= \sum_{T_f(i) \in \mathcal{T}_f(i)} \Pr\{T_f(i)\} \sum_{s \in T_f(i)} \frac{p(s)}{\Pr\{T_f(i)\}} E[X_i | S = s] \\
&\geq \sum_{T_f(i) \in \mathcal{T}_f(i)} \Pr\{T_f(i)\} E[\text{ergodic sojourn time of trap } T_f(i)] \\
&= \sum_{T_f(i) \in \mathcal{T}_f(i)} \Pr\{T_f(i)\} T_V(T_f(i))
\end{aligned}$$

(E2)

The first "$\geq$" comes from the fact that conditioned on any state $s \in G$, $E[X_i | S = s] \geq 0$. As for the second "$\geq$", consider the following. Since $T_f(i)$ is a frozen trap of link $i$, given that the process is already within $T_f(i)$, the residual waiting time of link $i$ must be larger than the ergodic sojourn time of the trap (i.e., when the system leaves the trap, link $i$ will still not be transmitting).

*3) Sufficient condition to identify starvation defined by (E1)*

If there exists a frozen trap $T_f(i) \in \mathcal{T}_f(i)$ satisfying

$$\Pr\{T_f(i)\} T_V(T_f(i)) \geq \Theta(\rho), \quad (E3)$$

then link $i$ will suffer from starvation for sufficiently large $\rho$. In this case, there exists a large enough $\rho$ such that

$$E[X_i] > \Pr\{T_f(i)\} T_V(T_f(i)) > \overline{X}_{\text{target}}.$$

Consider a two-links network in which the two links can hear each other. In equilibrium, each link has a normalized throughput of 0.5. It is easy to verity that there is no trap in this network. Hence, the links will not be starved according to the united definition.

In comparison, consider the network in Fig. 1 (a). In equilibrium, each link obtains a normalized throughput of 0.5. Now, $Tr_2 = G_2^{(2)} = \{\{2,3\},\{2,6\},\{3,6\},\{2,3,6\}\}$ is a frozen trap of links 1, 4 and 5. Based on the trap theory developed in Section V, we have $\Pr\{Tr_2\} T_V(Tr_2) = \Theta(\rho)$. That is, when $\rho$ is large, links 1, 4 and 5 will have an excessively long expected residual waiting time, then suffer from starvation. Similar analysis can be applied to links 2, 3 and 6. We conclude that links in this network suffer from starvation with respect to the unified definition of (E1).

Consider a circular network with five links in Fig. E1 (a). Its state-transition diagram is shown in Fig. E1 (b). There is no frozen trap for any link in the network since all the states are connected in column 1. Therefore, there is no trap in which any link will starve. Indeed, we know that the equilibrium throughput distribution is (0.4, 0.4, 0.4, 0.4, 0.4) and the temporal throughputs of each link observed are quite stable in the network.

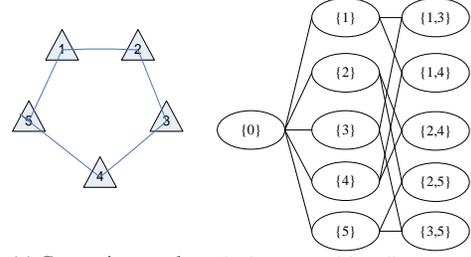

(a) Contention graph  (b) State-transition diagram
Fig. E1.  An example network without starvation.